\documentclass{aa}

\usepackage{epsfig,graphicx,lscape}
\begin{document}
\title{The VLT-FLAMES survey of massive stars: atmospheric parameters and
rotational velocity distributions for
B-type stars in the Magellanic Clouds\thanks{Based on observations at the 
European Southern
Observatory in programmes 171.0237 and 073.0234}} 

\author{I. Hunter\inst{1,2} \and D.J. Lennon\inst{2,3} \and P.L. Dufton\inst{1} 
  \and C. Trundle\inst{1} \and S. Sim\'{o}n-D\'{i}az\inst{4}
  \and S.J. Smartt\inst{1} \and R.S.I. Ryans\inst{1} \and C.J. Evans\inst{5} 
}

\institute{Astrophysics Research Centre, School of Mathematics \& Physics, The
   Queen's University of Belfast, Belfast, BT7 1NN, Northern Ireland, UK 
   \and The Isaac Newton Group of Telescopes, Apartado de Correos 321, 
   E-38700, Santa Cruz de La Palma, Canary Islands, Spain
   \and Instituto de Astrof\'{i}sica de Canarias, 38200 La Laguna, Tenerife, 
   Spain
   \and LUTH, Observatoire de Meudon, 5 Place Jules Janssen, 92195, Meudon
   Cedex, France
   \and UK Astronomy Technology Centre, Royal Observatory, Blackford Hill, 
   Edinburgh, EH9 3HJ
}

\offprints{I. Hunter,\\ \email{I.Hunter@qub.ac.uk}}

\date{Received; accepted }

\abstract{}
{We aim to provide atmospheric parameters and rotational velocities of a large
sample of O- and early B-type stars, analysed in a homogeneous and consistent
manner, for use in constraining theoretical models.}
{Atmospheric parameters, stellar masses and rotational velocities have 
been estimated
for approximately 250 early B-type stars in the Large (LMC) and Small
(SMC) Magellanic Clouds from high-resolution VLT-FLAMES data using the
non-LTE TLUSTY model atmosphere code.  This data set has been
supplemented with our previous analyses of some 50 O-type stars (Mokiem et al.
2006, 2007)
and 100 narrow-lined early B-type 
stars (Hunter et al. 2006, Trundle et al. 2007) from
the same survey, providing a sample of $\sim$400
early-type objects.}
{Comparison of the rotational velocities
with evolutionary tracks suggest that the end of core hydrogen burning occurs
later than currently predicted.
 We also show that the large 
number of the luminous blue supergiants observed in the fields
are unlikely to have directly evolved from main-sequence massive 
O-type stars as neither their low
rotational velocities or position on the
H-R diagram are predicted. We suggest that blue-loops or mass-transfer binary
systems may populate the blue supergiant regime.
By comparing the rotational velocity distributions of the Magellanic
Cloud stars to a similar Galactic sample we find that (at 3$\sigma$
confidence level) massive stars (above 8M$_{\rm \sun}$)
in the SMC rotate faster than those in
the solar neighbourhood.  However there appears to be no significant
difference between the rotational velocity distributions in the Galaxy
and the LMC. We find that the $v \sin i$ distributions in the SMC and 
LMC can modelled with an intrinsic rotational velocity distribution
which is a Gaussian peaking at  
175\,km\,s$^{-1}$ (SMC) and 100\,km\,s$^{-1}$ (LMC) 
with a $\frac{1}{e}$ half width of 150\,km\,s$^{-1}$. 
We find that in NGC346 in the SMC, the 10-25\,M$_{\rm \sun}$ 
main-sequence 
stars appear to rotate faster than their higher mass counterparts. 
It is not expected that O-type stars spin down significantly 
through angular momentum loss via stellar winds at SMC metallicity,
and hence this could be a reflection of mass dependent birth spin rates. 
Recently  Yoon et al. (2006) have determined rates of GRBs
by modelling rapidly rotating massive star progenitors. 
Our measured rotational velocity distribution for the 10-25M$_{\rm \sun}$
stars is peaked at slightly higher velocities
than they assume, supporting the idea that GRBs could come from 
rapid rotators with initial masses as low as 14\,M$_{\rm \sun}$
at low metallicities. }{}


\titlerunning{Rotational velocity distributions in the LMC and SMC}

\maketitle
%
\section{Introduction} \label{s_intro}

The evolution of massive stars has traditionally been described in
terms of mass and metallicity. However in the past 20 years it has
become evident that rotational effects are equally important, from
star formation through to their deaths in supernovae explosions and
associated gamma-ray bursts (Woosley \& Heger \cite{woo06}). For example, Heger \& Langer
(\cite{heg00}) discuss how rotation changes the lifetime and evolution
of massive stars on the Hertzsprung-Russell diagram through
rotationally induced mixing between the core and envelope.

Rotational effects are also believed to be metallicity
dependent. 
At low metallicity, line-driven winds are weaker 
and mass-loss rates lower (Kudritzki et al. \cite{kud87}; 
Kudritzki \& Puls \cite{kp2000};
Vink et al. \cite{vin01}; Mokiem et al. \cite{mok06}).
Hence a star will lose 
less angular momentum and thereby maintain
higher rotational velocities on the main-sequence. 
This promotes rotational mixing over a longer
period of an object's 
evolution which in turn extends its lifetime.
{\bf Additionally nucleosynthetically processed material will} be 
mixed from the core into the photosphere and hence abundance anomalies 
are expected, with 
helium and nitrogen enhancements and carbon and oxygen
depletions. Recent observations of massive stars 
have revealed abundance anomalies in O-stars (Bouret et al. \cite{bou03}, 
Mokiem et al. \cite{mok06}, \cite{mok07}), B-type supergiants (Dufton et al. 
\cite{duf05}, Hunter et al. \cite{hun06}, Lennon et al.
\cite{len03} and Trundle \& Lennon \cite{tru05}), B-type main-sequence stars
(Hunter et al. \cite{hun06}, Korn et al. \cite{kor02}, \cite{kor05}), 
A-type supergiants (Venn \cite{ven99}) and lower mass stars
(Smiljanic et al. \cite{smi06}).

Maeder \& Meynet (\cite{mae01}) have included rotation and metallicity effects
in stellar evolutionary models to explain both 
the relative frequency of blue and red supergiants as well as observed
abundance trends in A-type supergiants at low metallicity. However much remains
uncertain. For example, 
they adopt initial rotational velocities of 300\,km\,s$^{-1}$, which
equates to an average main-sequence rotational velocity of 220-240\,km\,s$^{-1}$
(there being little change in the rotational velocity over the main-sequence
lifetime of a B-type star), but statistically significant observational
samples are necessary both to estimate typical rotational velocities as well
as to calibrate the effect of metallicity on the rotational velocities. 
%
Additionally rotation has been
described as the key ingredient in enabling single stars to explode with
associated long-duration
gamma-ray bursts by Woosley \& Heger (\cite{woo06}), although rotational
velocities close to 400\,km\,s$^{-1}$ are required and
hence gamma-ray bursts would be
relatively rare events compared to supernovae. The requirement 
that little angular momentum
is lost during their stellar lifetime 
obviously favours low metallicity regimes with their smaller mass-loss rates.

A large
survey of massive stars has been undertaken at the European Southern
Observatory using the Fibre Large Array Multi-Element Spectrograph (FLAMES) on
the 8.2m Keuyen Very Large Telesope. The observations are discussed in detail in
Evans et al. (\cite{eva05}, hereafter Paper I) and Evans et al. (\cite{eva06},
hereafter Paper II) and are
summarised in Sect.~\ref{s_obs}. Briefly, some 750 stars
have been observed with mainly
O and early B spectral types in the Galaxy and the Large (LMC) and
Small (SMC) Magellanic Clouds. The analysis of many of these objects is well
underway or completed. 
Mokiem et al. (\cite{mok06}, \cite{mok07}) have derived atmospheric
parameters and helium abundances for the majority of the O-type stars in the
sample and discuss the dependence of mass-loss rates on metallicity and
calibrate the wind-momentum luminosity relation. Hunter et al. (\cite{hun06})
and Trundle et al. (\cite{tru07}) have derived atmospheric parameters and metal
abundances (CNO in particular) for a subset of the narrow-lined B-type
main-sequence and supergiant stars finding that significant mixing has occurred
in many of the objects. Dufton et al. (\cite{duf06a}, hereafter Paper III) have
estimated atmospheric parameters and rotational velocities of the entire
Galactic sample and report that more massive stars
have lower rotational velocities
(as expected from enhanced mass-loss) and that cluster objects appear to rotate
significantly faster than objects in the Galactic field.

In order to examine the effects of rotation, here
we perform a similar analysis to Paper III and estimate
atmospheric parameters and rotational velocities for the Magellanic Cloud sample
of B-type stars. {\bf The principle intention 
of performing such an analysis was to
provide rotational velocities for a large sample of early-type stars
in a homogenous and consistent manner. This dataset can then be used to answer
several questions of importance to massive star evolution:

\begin{itemize}
\item Where does the end of core hydrogen burning occur and how can large
populations of post core hydrogen burning objects be explained?
\item Are the rotational velocities of early B-type stars dependent
on the metallicity of the host environment?
\item Are mass-loss effects important at low metallicity
and how do these affect the predictions of gamma-ray burst rates from
massive star progenitors?
\end{itemize}
}


In Sect.~\ref{s_obs} we briefly describe the FLAMES survey and our selection
criteria for the objects discussed here. In Sect.~\ref{s_analysis} we present
our methods for deriving the atmospheric parameters and
projected rotational velocities. {\bf In Sect.~\ref{s_nature} the rotational
velocity distributions of the different subsets of the sample (for example,
binaries, Be-type stars, high mass stars and supergiants) are compared.
In Sect.~\ref{s_discussion} the relevant subsets are utilised to answer the
three questions posed above.} Finally in Sect.~\ref{s_conclude} we present the
principal conclusions from our analysis.

\section{Observations}                                             \label{s_obs}

As part of a European Southern Observatory (ESO) Large Programme on the 
Very Large Telescope (VLT, Paranal, Chile) observations were obtained for over
700 stars using the Fibre Large Array Multi-Element Spectrograph (FLAMES), with
the majority of these objects being O- or early B-type stars. Observations were
taken towards three Galactic clusters, NGC\,6611, NGC\,3293 and NGC\,4755; two
LMC clusters, N\,11 and NGC\,2004 and two SMC clusters, NGC\,346 and NGC\,330.
Details of the observational fields, target selection and data reduction were
discussed in Paper I and 
Paper II\footnote{We note that in Paper II the radial distance
of the NGC\,346 objects from the centre of the association was calculated
based on the NGC\,346 centre obtained from
the SIMBAD database (operated at the CDS, Strasbourg, France) which is actually 
the centre of the ionised shell of gas and is offset from the centre of the
OB association. In Table~\ref{t_346_rad} (only available online at the CDS)
the radial distances are listed
with the centre taken as the position of object 435 
from the Massey et al. (\cite{mas89}) catalogue of OB stars in NGC\,346;
$\alpha$(2000)\,=\,00\,59\,04.49 $\delta$(2000)\,=\,-72\,10\,24.7.}.
The FLAMES
spectroscopy covered the wavelength regions 3850-4755\AA\ and 6380-6620\AA\ and
was supplemented with additional
observations
from the Fiber-fed Extended Range Optical Spectrograph (FEROS) and the
Ultraviolet and Visual Echelle Spectrograph (UVES), both of which had a more 
extensive
wavelength coverage. To maintain consistency throughout the analysis we
only consider the 
FEROS and UVES data in the same wavelength range as that of the 
FLAMES observations.

\subsection{Radial velocity corrections and binarity}         \label{s_rad_corr}

Observations were taken in six wavelength settings in order to 
achieve our desired wavelength coverage at
high spectral resolution, 
with multiple exposures (typically six) being taken in
each setting. {\bf However, in cases where 
exposures are of significantly lower quality than the norm,
these have been excluded}. 

To increase the signal to noise (S/N) ratio of the data, it is possible to
combine the individual exposures in each setting. However, it is first necessary
to check for radial velocity shifts between
exposures, an indicator of binarity. {\sc IDL} procedures have been developed to
cross-correlate each exposure in the FLAMES data and identify significant radial
velocity shifts at the 3$\sigma$ level (see Hunter et al. \cite{hun06}).
We note that
this procedure is superior to that used in Papers I and II where binaries were
identified by eye from the combined spectra of the exposures in each wavelength
setting. Our cross-correlation procedure has identified 
additional binaries not previously
detected in Paper II. Additionally for several stars
we find that the radial
velocity shifts detected in Paper II are not significant at the 3$\sigma$ level
and hence we do not consider them as binary stars. We note that most of
these objects were identified as possible binaries in Paper II or had Be
classifications and large rotational velocities, which makes binary detection
uncertain. After correction
for velocity shifts, 
{\sc IRAF} routines were used to combine the exposures
and simultaneously remove cosmic rays. This procedure was not possible for the
double lined spectroscopic binaries and the analysis of these systems
(where this was possible) is discussed in Sect.~\ref{s_vsini_SB2}. Throughout
this paper all radial velocity variables are considered as binary objects
although our
temporal resolution is not sufficient to derive periods and other orbital
parameters to confirm
if they truly are all binary stars.

\subsection{Selection criteria}

We have selected objects with spectral types later than O\,9 and earlier than
B\,9 which can be analysed with our
non-LTE {\sc TLUSTY} model atmosphere grid which
has an effective temperature range from 12\,000\,K to 35\,000\,K
(see Sect.~\ref{s_tlusty}). The majority of the O-type stellar sample
observed in the FLAMES survey
has been analysed by Mokiem et al. (\cite{mok06}, \cite{mok07})
for the SMC and LMC O-type stars respectively.

We have attempted to derive parameters for every object in our spectral type
range but
this was not possible for a small minority.
Typically objects were excluded if it was not possible to separate
individual spectra in a double lined binary system, if the spectra were not of
sufficient quality to reliably estimate velocity corrections in the case of
single lined spectroscopic binaries or if the stellar spectra appeared to be
variable between exposures. Additionally a few objects
have been excluded as shell absorption
contaminated both the hydrogen and helium lines and no metal lines were 
well enough observed
to allow for a reliable determination of the rotational velocity. In
Table~\ref{t_exclude} we list those objects within our spectral type range that
have been excluded. These objects form less than 6\% of
the sample, with the majority being binaries and having
either Be classifications or uncertain spectral types.

\begin{table}
\addtocounter{table}{1}
\centering
\caption[]{Objects with spectral types later than O\,9 and earlier than B\,9
observed in the FLAMES survey that have been excluded from this analysis.}
\label{t_exclude}
\begin{tabular}{ll}
\hline \hline
Cluster  & \multicolumn{1}{c}{Excluded stars} \\
\hline \\
NGC\,346 & 017, 038, 060, 086, 105, 111, 115 \\
NGC\,330 & 030, 077, 088, 092, 093, 115, 117 \\
N\,11    & 005, 030, 053, 067, 099, 112      \\
NGC\,2004& 019, 028, 033, 034, 037, 038, 072 \\
\\ \hline
\end{tabular}
\end{table}

\subsection{{\bf Field populations}}\label{s_clustermem}

In Paper II images of the four regions are displayed and it is clear that the
majority of our objects do not lie within the core of the clusters,
due to constraints in positioning the fibres of the FLAMES instrument.
Specifically for the NGC\,330 and NGC\,2004 fields, our samples are
probably dominated
by field objects. NGC\,346 does not have a compact core and it is difficult to
determine the extent of the cluster. Indeed, NGC\,346 is often 
described as an OB association rather than as a bound cluster (see for example
Gouliermis et al. \cite{gou06}). 
N\,11 is
dominated by two distinct regions, LH\,9
and LH\,10, where sequential star formation has occurred. Additionally there are
many smaller pockets of star formation within this region complicating the
picture further. 

The supergiant
objects in NGC\,330 are not located close to or in the core of NGC\,330; they
instead appear to be randomly distributed in the field and may indicate that
they are indeed field stars. However, it should be noted that 
almost no supergiants are observed
towards NGC\,346, implying an age difference between NGC\,346 and NGC\,330.
Similarly to NGC\,330 the 
supergiants observed towards N\,11 and NGC\,2004 
are not clustered in one region. 
Additionally in N\,11 the O-type stars are not clustered
in the younger LH\,10 region, indicating that the OB association is probably not
bound. 

As such our four Magellanic Cloud samples are probably
better described as unbound associations with the association
`evaporating' as their OB stars
become part of the field population. Sirianni et al. (\cite{sir02}) have
obtained colours for the stars in both the central and outer regions of
NGC\,330 and find remarkable similarity between the regions implying that
NGC\,330 may extend well beyond its dense core. 
The recent review of Bastian \& Gieles (\cite{bas06})
describes the various phases of cluster disruption, in particular
that of `infant mortality', where through gas expulsion during cluster formation
the cluster becomes unbound, and if the star formation efficiency is low, the
cluster can become unbound on the order of a few tens of Myr 
(see the recent work
of Goodwin \& Bastian \cite{goo06} and references therein).

\section{Analysis}                                            \label{s_analysis}

\subsection{Non-LTE {\sc tlusty} model atmosphere grids}        \label{s_tlusty}

The non-LTE {\sc tlusty} model atmosphere grids (see Hubeny \& 
Lanz \cite{hub95} and references therein) used in this analysis have
previously been discussed -- see Hunter et al. (\cite{hun06}) for an overview
of the grids and Dufton et al. (\cite{duf05}) for a detailed
discussion\footnote{See also http://star.pst.qub.ac.uk}.

Briefly, model atmosphere grids have been calculated for the analysis of B-type
stellar spectra, covering the effective temperature range from 12\,000\,K to
35\,000\,K in steps of not more than 1\,500\,K, a surface gravity range from
the Eddington limit up to 4.5\,dex, in steps of 0.15\,dex and
microturbulence values of 0, 5, 10, 20 and 30\,km\,s$^{-1}$. Four model
atmosphere grids have been generated with metallicities representative of the
Galaxy ([Fe/H]=7.5\,dex), the LMC (7.2\,dex), the SMC (6.8\,dex) and a lower
metallicity regime (6.4\,dex). A nominal helium abundance of 11.0\,dex has been
adopted throughout. Theoretical spectra and metal line equivalent widths 
have then been calculated for each grid point.

\subsection{Atmospheric parameters}                         \label{s_parameters}

Hydrogen and helium lines are observable in the spectra of all our 
targets and these lines have been used as the primary
indicators of effective temperature and surface gravity.
IDL procedures have been developed which 
fit model spectra (convolved with an appropriate stellar broadening
function) of the appropriate metallicity 
to the observed hydrogen and helium lines and calculate the region 
of best
fit (by chi-squared minimisation) in effective temperature - surface gravity
space. As these parameters are interdependent it is necessary to use an 
iterative method. In
Paper III equivalent width measurements of the hydrogen and helium lines were
used to constrain the atmospheric parameters but we believe that the `profile
fitting' methodology we have adopted here is an improvement 
as it uses both information on the line strength and shape. We estimate that
this methodology results in a typical uncertainty of 1\,500\,K for the effective
temperature, with that for the surface gravity being discussed below.

The \ion{He}{i} spectrum is a suitable effective
temperature indicator at spectral types
of approximately B\,3 and later, while the \ion{He}{ii} spectrum can be
used to constrain the temperature of stars with a spectral type of B\,1 or
earlier. Hence there remains a range of spectral types where the
\ion{He}{ii} lines
are not observed and the \ion{He}{i} spectra are not sensitive to temperature.
For spectra with narrow metal absorption lines,
where two ionization stages of silicon were
observed, it is possible to use the silicon ionization balance to constrain the
effective temperature, with the surface gravity again being
estimated from the hydrogen
lines. Hunter et al. (\cite{hun06}) and Trundle et al. (\cite{tru07}) have 
presented 
analyses of the narrow lined B-type stars in the FLAMES survey and
their atmospheric parameters have been adopted. 
For stars with large projected rotational velocities in the B\,1-3 spectral type
range, it was not possible to observe
two stages of silicon ionization. As in Paper III, we have 
adopted effective
temperature-spectral type calibrations. These are based on 
the analysis of the narrow lined objects and are presented in Trundle et
al. for both LMC and SMC metallicities.

Both the H$\delta$ and H$\gamma$ absorption
lines have been observed in our spectra and these have been used
to estimate the surface gravity by the method of profile fitting described 
above. 
In the majority of cases we find excellent agreement between the estimate
derived from each line with differences typically being less that
0.15\,dex. The effective temperature and surface gravity estimates are
directly correlated and an over estimate of the effective temperature by
1\,000\,K will
lead to an over estimate of the surface gravity by approximately 0.1\,dex. Hence
we believe that our surface gravity estimates should be typically accurate to
0.2-0.25\,dex.

It has previously been found that there is a correlation 
between surface
gravity and microturbulence with supergiants typically having greater values of
microturbulence than giant and main-sequence stars (see for example Gies \&
Lambert \cite{gie92}).
Representative 
microturbulence values of 10 and 5\,km\,s$^{-1}$ have been adopted for
supergiants and the less evolved objects respectively. It should be
noted that this choice has a negligible effect on the atmospheric
parameters and only affects the projected rotational velocity estimate when the
measured values are less than 20\,km\,s$^{-1}$. The atmospheric parameters are
listed in Tables~\ref{t_346_vsini}-\ref{t_2004_vsini}, only available online at
the CDS. In these tables we give the star identifier and spectral type (from
Paper II), the effective temperature, surface gravity, projected rotational
velocity, adopted methodology, luminosity, mass and additionally indicate if the
star is a radial velocity variable and hence probably in a binary system.

\subsection{Projected rotational velocities}              \label{s_vsini_method}

Several methods are available to estimate projected rotational velocities
($v\sin i$). The most commonly used
in the case of OB-type stars are based on a direct measure
of the FWHM of the spectral lines (viz. Slettebak et al. \cite{sle75};
Abt et al. \cite{abt02}; Herrero et al. \cite{her92}; Strom et al. \cite{str05})
or profile fitting (Gray \cite{gra76}; Paper III), while
Penny (\cite{pen96}) and Howarth et al. (\cite{how97}) applied a cross 
correlation method to IUE spectra. 
Recently, Sim\'{o}n-D\'{i}az \& Herrero (\cite{sim06}) have discussed the
utility of the Fourier method (Gray \cite{gra76})
in the case of early type stars. Each of these methods 
have their advantages and disadvantages, with for
example, the Fourier method being able to deconvolve the rotational
broadening from other broadening mechanisms such as macroturbulence, although it
requires
high quality spectra. Profile
fitting to obtain the rotational velocity is possible for lower quality
spectra, but relies on an appropriate description of the intrinsic line
profile (from a stellar atmosphere code), and the
broadening agents affecting the line profile. Measuring
the FWHM and using the cross correlating method is 
relatively straight forward but does not take into account the possible extra
non-rotational broadenings (such as Stark broadening of the 
H and He lines, macroturbulence and microturbulence) thereby 
leading in some cases to 
only upper limits for the projected rotational velocity estimate.

Given the quality of our spectra we decided to use the profile fitting
methodology discussed in Paper III for all the non-supergiant objects as the
quality
of our data is not high enough to make the Fourier method viable for the
entire sample.
Briefly, a model 
spectrum, at the closest {\sc tlusty} grid point
to the derived atmospheric parameters, was selected and the equivalent width of
an appropriate line
was then scaled to the same strength as that observed. Typically this
scaling was less than 10\% and, except for the narrowest lined spectra, has
little effect on the derived projected rotational velocity. The model profile 
was first convolved with a
Gaussian profile to take into account the spectral resolution of our data; 
then, the resultant profile was convolved with a rotational broadening function
for a range of rotational velocities. A
chi-squared minimisation test was used to estimate the projected
rotational velocity. The
atmospheric parameters were then redetermined with this new estimate
and the process repeated. Generally no more than two iterations were 
required.

Hydrogen lines were 
observed in all our spectra, but their profiles are dominated
by Stark broadening. Although this also affects the diffuse \ion{He}{i} 
lines, at 
moderate projected rotational velocities, the rotational broadening begins to
dominate, and
hence they are suitable for
estimating this quantity. The \ion{He}{i} 4026\AA\ line is observed
across all our spectral types 
and additionally is well separated from other absorption
features. Therefore this line was employed for the profile fitting
methodology. Note however, that at low projected rotational velocities the 
profile of this \ion{He}{i} line is no longer
sensitive to the rotational broadening.  For these cases, 
we have instead used metal lines if
available
(typically either the \ion{Mg}{ii} 4481\AA\ line or the \ion{Si}{iii}
4552\AA\ line).
These projected rotational velocity estimates
are listed in Tables~\ref{t_346_vsini}-\ref{t_2004_vsini}, apart from those of
the supergiant targets discussed below.

A small fraction of the objects in our sample have uncertain spectral types
that lie in the range B\,1-3 and, as discussed above, 
the He line spectrum is not
capable of constraining their temperatures. 
In order to derive the projected rotational velocity of these
objects, approximate models have been adopted.
As these stars frequently have large
rotational velocities (leading to the uncertain spectral classification), 
the broadening of the observed lines will
be rotationally dominated and hence assuming an inappropriate model
will not significantly affect the derived rotational velocity.
Atmospheric parameters  are not given in
Tables~\ref{t_346_vsini}-\ref{t_2004_vsini} for these objects.

\begin{figure}[ht]
\centering
\epsfig{file=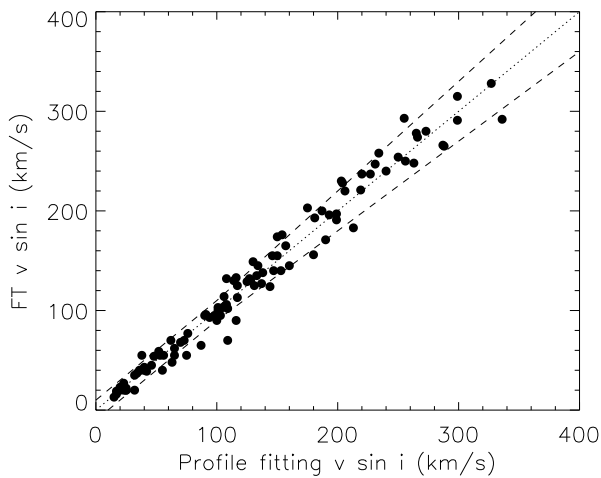, height=70mm, angle=0}
\caption[]{Comparison of the projected rotational velocity derived by the
profile fitting method and the Fourier transform method for the non-supergiant
objects in the sample. 
The dotted line indicates a one-to-one correlation. The dashed
lines indicate a 10\% or 10\,km\,s$^{-1}$ uncertainty, whichever is the larger.}
\label{f_FT_plot}
\end{figure}

We have also estimated rotational velocities using the Fourier transform
methodology where possible. A detailed discussion of the applicability of this 
methodology in the case of early type stars
is given by Sim\'{o}n-D\'{i}az \& Herrero (\cite{sim06})
and will not be repeated here. In
Fig.~\ref{f_FT_plot} we compare the projected rotational velocity derived by the
profile fitting methodology and the Fourier transform methodology for the
non-supergiant targets, where the spectra were of sufficient quality for the
application of the Fourier
method. We find
in almost all cases good agreement (within 10\%),
indicating that the choice of methodology is unimportant
for these targets and validating our adopted projected rotational
velocity estimates.

\begin{figure}[ht]
\centering
\epsfig{file=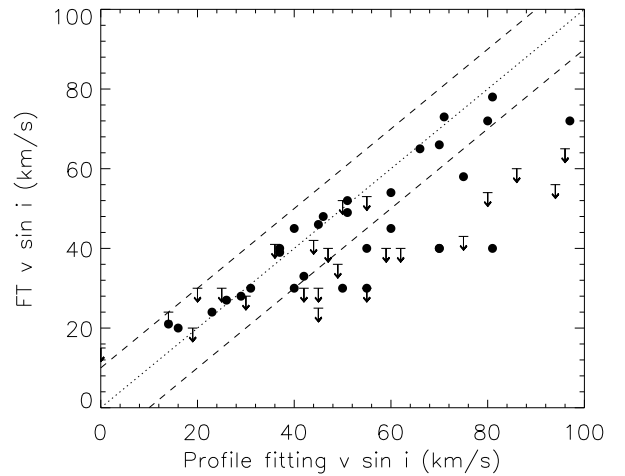, height=70mm, angle=0}
\caption[]{Comparison of the projected rotational velocity ($v\sin i$) 
derived by the
profile fitting method and the Fourier transform method for the supergiant
objects in the sample. 
The dotted line indicates a one-to-one correlation. The dashed
lines indicate a 10\,km\,s$^{-1}$ uncertainty. Downward
pointing arrows
indicate those objects where it was only possible to derive upper
limits to the $v\sin i$ from the Fourier method.}
\label{f_FT_SG}
\end{figure}

As already proposed by several authors (Slettebak \cite{sle56}; Conti \& Ebbets
\cite{con77}; Howarth et al. \cite{how97}), and recently illustrated by Ryans
et al. (\cite{rya02}), Dufton et al. (\cite{duf06b}) and 
Sim\'{o}n-D\'{i}az \& Herrero (\cite{sim06}), there is an important
non-rotational broadening mechanism affecting the profiles of early type
supergiants (usually termed macroturbulence). Since the line fitting
methodology does not allow for this extra source of broadening,
the Fourier transform technique has been used 
to derive the projected rotational velocity of objects
classed as supergiants in the sample. In Fig.~\ref{f_FT_SG} we compare 
the derived
rotational velocities from the two methods for these objects. 
Due to the quality of the spectra, in terms of signal to noise ratio, in many
cases it was only possible to derive upper limits to the projected rotational
velocity from the Fourier method
(see Sim\'{o}n-D\'{i}az \& Herrero); these 
are indicated by downward pointing arrows in Fig.~\ref{f_FT_SG}.

From Fig.~\ref{f_FT_SG} it is apparent that in many cases the projected
rotational velocities derived from the Fourier Transform method are 
significantly 
lower than those from the profile fitting method and these are likely to be
objects with significant macroturbulent broadening. Hence for the supergiant
targets, the projected
rotational velocities listed in
Tables~\ref{t_346_vsini}-\ref{t_2004_vsini} are those deduced from the Fourier
Transform methodology. We note that for the determination of the atmospheric 
parameters of the supergiant
objects in the sample, the estimated rotational velocity 
determined from the profile fitting method
was used rather than that from the Fourier Transform method. This is valid as it
is the entire broadening of the line which is important when determining the
parameters and not simply the rotational component.

The uncertainties in effective temperature and surface gravity have
little effect upon our estimated projected rotational velocities. 
For example, 
NGC\,346-083 was analysed with a model profile at a temperature
of 27\,500\,K 
and a surface gravity of 4.00\,dex and a projected rotational velocity of
207\,km\,s$^{-1}$ was estimated. Adopting a higher temperature
profile of 30\,000\,K and higher surface gravity of 4.25\,dex (as the surface
gravity and effective
temperature estimates 
are directly correlated) leads to a $v\sin i$ estimate of only
4\,km\,s$^{-1}$ greater. Similarly adopting a lower temperature and gravity
(25000\,K, 3.75\,dex) results in a value 14\,km\,s$^{-1}$ greater. 
Including the effects of normalisation, scaling and choice of model profile we
would not
expect the errors in our rotational velocity estimates to exceed 10\%. This is
confirmed by {\bf the comparison shown in} 
Fig.~\ref{f_FT_plot} where different lines were used for
the Fourier Transform and the profile fitting methods. We note that the Fourier
Transform method is additionally model independent.

Rapidly rotating objects (close to critical velocity) can not be considered
spherical and equatorial gravity darkening (the cooler and fainter
equatorial regions of an oblate star
contribute less flux to the observed spectrum) can
significantly alter the estimated atmospheric parameters and projected rotational
velocities (von Zeipel \cite{von24}). 
Townsend et al. (\cite{tow04}) have shown that this
effect leads to an underestimation of projected rotational velocities for close
to critical rotational velocities and that the maximum ratio of projected
rotational velocity to critical velocity ($v\sin i$/$v_{\rm c}$)
that can be inferred from line widths is
approximately 0.8. We have derived critical rotational velocities for our sample
and find that {\bf less than 5\% of our sample} have $v\sin
i$/$v_{\rm c}$ ratios greater than 0.5 with the maximum {\bf value of this
ratio} being approximately 0.7
(NGC\,2004-113). If we consider that this object is rotating at near to critical
velocity, the projected rotational velocity would be
underestimated by less than 15\% due to the effects of gravity darkening
(Fig.~1, Townsend et al.).
Hence we 
believe that gravity darkening effects should be small
compared to our
observational uncertainties for the majority of 
our sample of stars and in the most extreme cases of the same order.

\subsubsection{{\bf Comparison with other work}} \label{s_mar_compare}

\begin{figure}[ht]
\centering
\begin{tabular}{c}
\epsfig{file=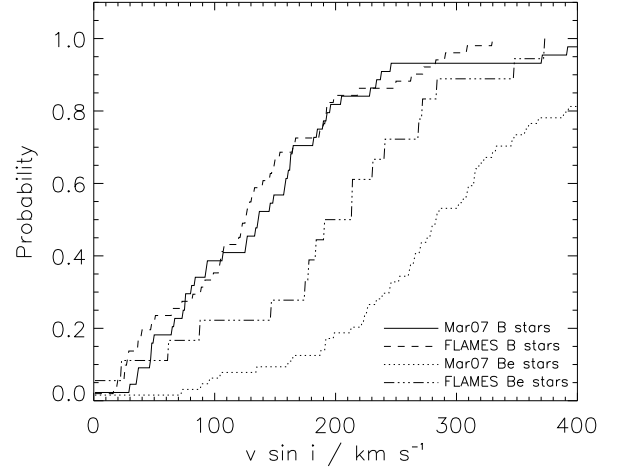, height=70mm, angle=0} \\
\end{tabular}
\caption[]{Comparison of our estimated rotational velocities (FLAMES) 
with those
of Martayan et al. (\cite{mar06b}, Mar07) for the B and Be type stars towards
NGC\,330. {\bf This clearly shows the
discrepancy between the Be analysis presented here and that of Martayan et al.}
Only objects in the visual magnitude range 15-16.5 have been included.
}
\label{f_marcomp}
\end{figure}

\begin{figure}[ht]
\centering
\begin{tabular}{c}
\epsfig{file=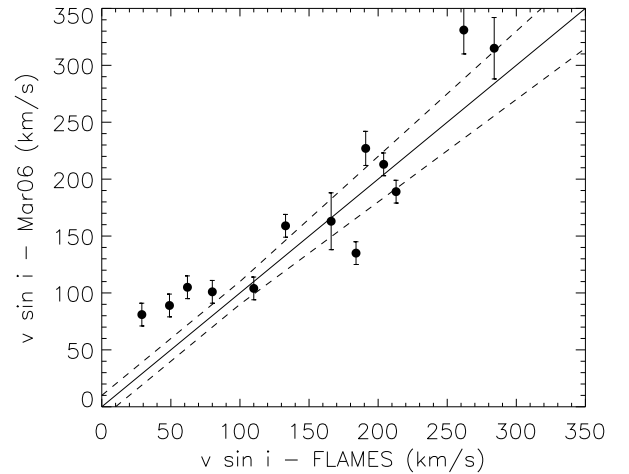, height=70mm, angle=0} \\
\end{tabular}
\caption[]{Comparison of our estimated rotational velocities (FLAMES) 
with those
of Martayan et al. (\cite{mar06b}, Mar06). The solid 
line indicates a one-to-one correlation. The dashed
lines indicate our 10\% uncertainty down to a limit of 10\,km\,s$^{-1}$.}
\label{f_vsini_comp}
\end{figure}

Martayan et al. (\cite{mar06a}, \cite{mar06b}) 
have observed a large number of Be and B-type stars towards
NGC\,2004 and NGC\,330 also with the FLAMES instrument, but
with lower spectral resolution than we employed. 
 In order to make comparison
with the results of 
Martayan et al. the samples were constrained to the magnitude
range where there is a good overlap between them. For NGC\,330 this
is the visual magnitude range of 15.0-16.5 {\bf with a comparison of the two
samples being shown in Fig.~\ref{f_marcomp}}. 
The cumulative probability distributions of the B-type
stars closely agree. There is some suggestion that the FLAMES sample
contains a larger fraction of stars with 
very low projected rotational velocities than does the that of Martayan et
al. Given the higher spectral resolution and signal to noise ratio 
of our FLAMES data, this is not unexpected.
There are 13 objects in common between our FLAMES survey of
NGC\,330 and that of Martayan et al. and in Fig.~\ref{f_vsini_comp} we compare 
the projected rotational velocities for these stars. 
In general there is reasonable agreement
between the two samples, although the discrepancy at low projected 
rotational velocities is again evident.

\begin{figure}[ht]
\centering
\epsfig{file=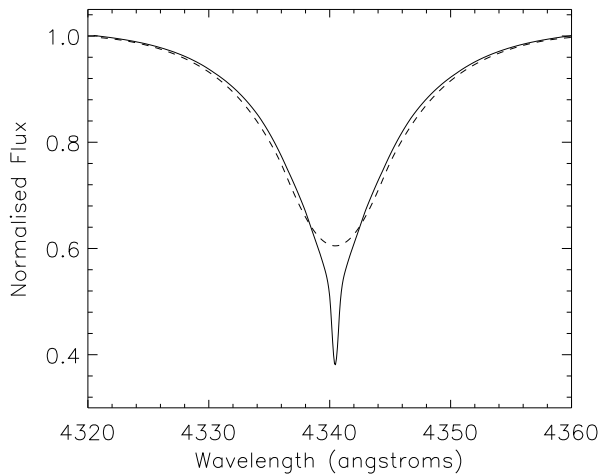, height=70mm, angle=0}
\caption[]{Comparison of a theoretical H$\gamma$ line with no rotational
broadening (solid line) with the same line rotationally broadened by
300\,km\,s$^{-1}$ (dashed line) {\bf showing that the wings of the hydrogen
line are not sensitive to rotational broadening.} }
\label{f_hlines}
\end{figure}

The distributions of the Be stars shown in Fig.~\ref{f_marcomp}
are in clear contrast, with Martayan et al. (\cite{mar06b}) finding
systematically higher
projected rotational velocities. 
The reason for such a large discrepancy is unclear. For the
estimation of parameters of their Be sample, Martayan et al. removed the
core of the lines where Be emission was observed and hence the
projected rotational velocities
are estimated from the wings of the line. In Fig.~\ref{f_hlines} we present
a comparison between a model spectrum of a H$\gamma$ line with no rotational
broadening and that with a rotational broadening of 300\,km\,s$^{-1}$. It is
clear that the
wings are not as sensitive as the line core to changes in the rotational
broadening.  
As such removing the line core will increase the uncertainty of the rotational
velocity estimate, especially for low
signal to noise data. However, by normally using only one line to
determine the projected rotational velocity, our estimates may be biased towards
smaller rotational velocities due to saturation effects (Fremat et al.
\cite{fre05}) compared to those of Martayan et al. who have used several lines,
although, this should only be an issue at high rotational velocity. As
discussed {\bf above} gravity darkening should be
negligible for the
objects in our sample. Indeed, for {\bf ten} Be-type objects we have also
derived rotational
velocities from the \ion{He}{i} line at 4387\AA\ and generally find good
agreement with the values from the 4026\AA\ line. 
Additionally in estimating the projected rotational
velocities, models with a surface gravity of 4.0\,dex have been adopted. At this
high gravity the Stark broadening will be relatively large
and hence less rotational
broadening is required to fit the profile compared to assuming a lower gravity,
see Table~\ref{t_Be_profiles}. As such our Be star rotational velocities may be
biased towards lower values, but this is only an issue at low projected
rotational velocities and in these cases we have endeavored to use metal lines
where possible in order to minimize this bias.

We have no NGC\,2004 objects in common with the sample of Martayan et al.
(\cite{mar06a}) and the magnitude range of overlap is much smaller than that of
the NGC\,330 samples although similar trends are seen in the rotational
velocity distributions. The B-type stars are in excellent agreement and again
there is some evidence that the Be stars of Martayan et al. have higher
rotational velocities. Although the discrepancy between the Be star analyses
cannot be fully resolved, this does not invalidate the metallicity dependence
discussed {\bf in Sect.~\ref{s_vmet}. Indeed, since} 
the SMC has a higher fraction of Be stars than the LMC,
using the results of Martayan et al. would enhance the difference
between the rotational velocity distributions of the two Magellanic Clouds. 

\subsubsection{{\bf Selection effects}}

As discussed in Paper II approximately 30 (potential) Be stars were removed from
the list of possible targets for the FLAMES pointing towards NGC\,330.  Given 
that Be stars are typically the fastest rotators in the sample, this could 
bias our velocity distribution towards lower rotational velocities.  To 
investigate this we ran the target selection software,
{\sc configure}, again for the NGC\,330 field, without 
explicitly excluding any objects from the input catalogue. From this test we 
found seven of the previously excluded stars were included in the resulting 
fibre configuration (which of course differs from our actual observed field 
by more than seven stars, due to the different positions of the allocated 
fibres).  If we consider that the net difference was to replace seven normal 
B-type stars with seven Be targets, the mean rotational velocity of the 
SMC stars would only increase by $\sim$3\,km\,s$^{-1}$.  Given that some of 
the excluded objects are not actually confirmed Be stars this is a 
conservative estimate, and we believe that any minor selection effect 
that may have been introduced does not unduly bias our conclusions.

\subsection{O-type stars}                                      \label{s_O_stars}

Atmospheric parameters and projected rotational velocities of the majority 
of the O-type
objects observed Paper II have been derived by Mokiem et
al. (\cite{mok06}, \cite{mok07}) and these have been directly inserted into 
Tables~\ref{t_346_vsini}-\ref{t_2004_vsini}.
We note that for those stars where our analysis overlaps with that of Mokiem et
al. we find good agreement between the two sets
of atmospheric parameters and rotational velocities. 

\subsection{Binary objects}                                  \label{s_vsini_SB2}

\begin{figure}[ht]
\centering
\epsfig{file=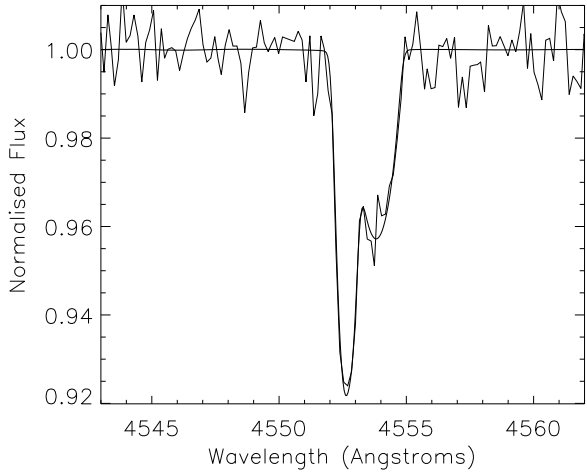, height=70mm, angle=0}
\caption[]{Fit of the \ion{Si}{iii} line at 4552\AA\ for the double lined
spectroscopic binary system NGC\,346-020. The model spectra for the primary and
secondary  objects have been fitted
with a $v\sin i$ value of 34\,km\,s$^{-1}$ and 66\,km\,s$^{-1}$ 
respectively.}
\label{f_SB2_fit}
\end{figure}

After correcting for radial velocity shifts (see Sect.~\ref{s_rad_corr}), the
single lined spectroscopic binaries were analysed using the methodology outlined
above. However, this was not possible for the
double lined spectroscopic binaries, 
due to both continuum contamination and the double line nature of the 
profiles. In Tables~\ref{t_346_vsini}-\ref{t_2004_vsini} we indicate those
objects for which we see evidence of a secondary object, although for small
asymmetries and velocity shifts in the line profile,
variability may also be an explanation.
For all cases, the individual exposures were examined to find 
those in which the line profiles from the primary and secondary objects had the
maximum separation. The profile fitting methodology was then
used to estimate the
rotational velocity of the
primary, and where possible, the secondary object. We do not attempt to 
derive atmospheric parameters for either object.
Metal lines were used to estimate the rotational velocities in order to minimise
the uncertainty from the choice of model profile. An example of the fit 
for the double lined
spectroscopic binary system NGC\,346-020 is shown in Fig.~\ref{f_SB2_fit}, 
where the \ion{Si}{iii}
lines from each object are partially blended. Where it has been 
possible to estimate
the projected rotational velocity of the secondary object in a binary system
this is listed in Tables~\ref{t_346_vsini}-\ref{t_2004_vsini} with the
letter `B' appended to the system's identifier.

\subsection{Be objects}                                       \label{s_vsini_Be}

\begin{table*}[ht]
\addtocounter{table}{4}
\caption{Dependence of the derived $v\sin i$ value upon surface gravity for a
sample of Be objects. For large rotational velocities, the \ion{Si}{iii} lines
could not be identified.}
\label{t_Be_profiles}
\centering
\begin{tabular}{lcccc}\hline \hline
Line                   & $\log g$&\multicolumn{3}{c}{$v\sin i$ (km\,s$^{-1}$)}\\
                       & (dex)   &  NGC\,346-091 & NGC\,346-095 & NGC\,346-096\\
\hline
\ion{He}{i} 4026\AA\   & 4.00     &  74          & 227          & 346         \\
\ion{He}{i} 4026\AA\   & 3.50     & 120          & 250          & 343         \\
\ion{Si}{iii} 4552\AA\ & 4.00     &  49          &  -           &  -          \\
\ion{Si}{iii} 4552\AA\ & 3.50     &  47          &  -           &  -          \\
\\
\hline
\end{tabular}
\end{table*}

A significant fraction of the stellar sample is made up of Be type objects,
with the majority having spectral types in the range B\,1-2. In
Tables~\ref{t_346_vsini}-\ref{t_2004_vsini}, we do not give atmospheric
parameters for these objects due to problems associated with continuum 
contamination
from the circumstellar disc and emission lines. 
Additionally as many of the objects do not
have precise spectral types it was not possible to assign an effective
temperature based on spectral type. 
We have however used the metal line and helium 
spectra to estimate the projected rotational broadening.

In Table~\ref{t_Be_profiles} we show how the estimated {\bf projected 
rotational velocity}
depends on the adopted surface gravity for three Be type stars with 
low, medium and high velocities.
It is evident that at large rotational velocities the He profile is
dominated by the rotational broadening and is no longer sensitive to the adopted
profile. Additionally the projected rotational velocity estimated 
from the metal line spectra does not
appear to be dependent on the adopted model profile 
and we have adopted these
estimates in Tables~\ref{t_346_vsini}-\ref{t_2004_vsini} where possible.

\subsection{Luminosities} \label{s_lum}

Luminosities were calculated for each object in the sample using the same
methodology as discussed in Hunter et al. (\cite{hun06}). Constant values of
reddening towards each cluster were assumed, with $E(B-V)$ values of 0.09
(Massey et al. \cite{mas95}), 0.06 (Gonzalez \& Wallerstein \cite{gon99}),
0.13 (Massey et al.) and 0.09 (Sagar \& Richtler \cite{sag91}) 
being adopted for NGC\,346, NGC\,330, N\,11 and
NGC\,2004 respectively. We adopt the SMC reddening law of $A_{\rm
V}$\,=\,2.72$E(B-V)$  from Bouchet et al. (\cite{bou85}) for the SMC targets
while we use the standard Galactic law for the LMC targets. We note that
assuming the standard Galactic law would
change the derived SMC luminosities by less
than 0.04\,dex. Bolometric corrections have been adopted from Vacca et al.
(\cite{vac96}) {\bf for objects hotter than 28,000\,K} and Balona (\cite{bal94})
{\bf for cooler objects, together with distance} modulii of
18.91\,dex (Hilditch et al. \cite{hil05}) and 18.56\,dex (Gieren et al.
\cite{gie05}) for the SMC and LMC. The luminosities are given in
Tables~\ref{t_346_vsini}-\ref{t_2004_vsini} and typically have an uncertainty of
0.1\,dex.

\begin{figure*}[htbp]
\centering
\begin{tabular}{cc}
\epsfig{file=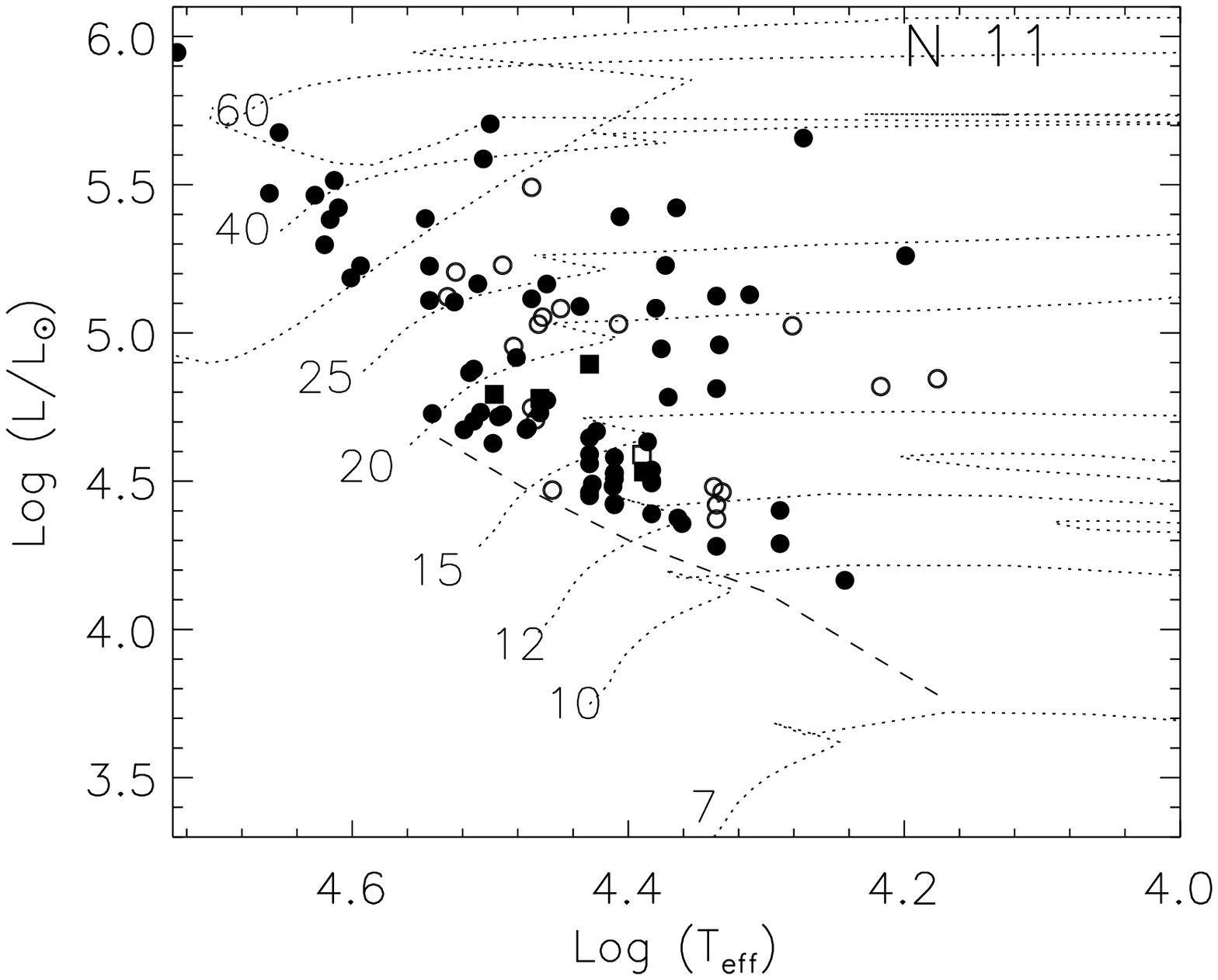, height=70mm, angle=0}&
\epsfig{file=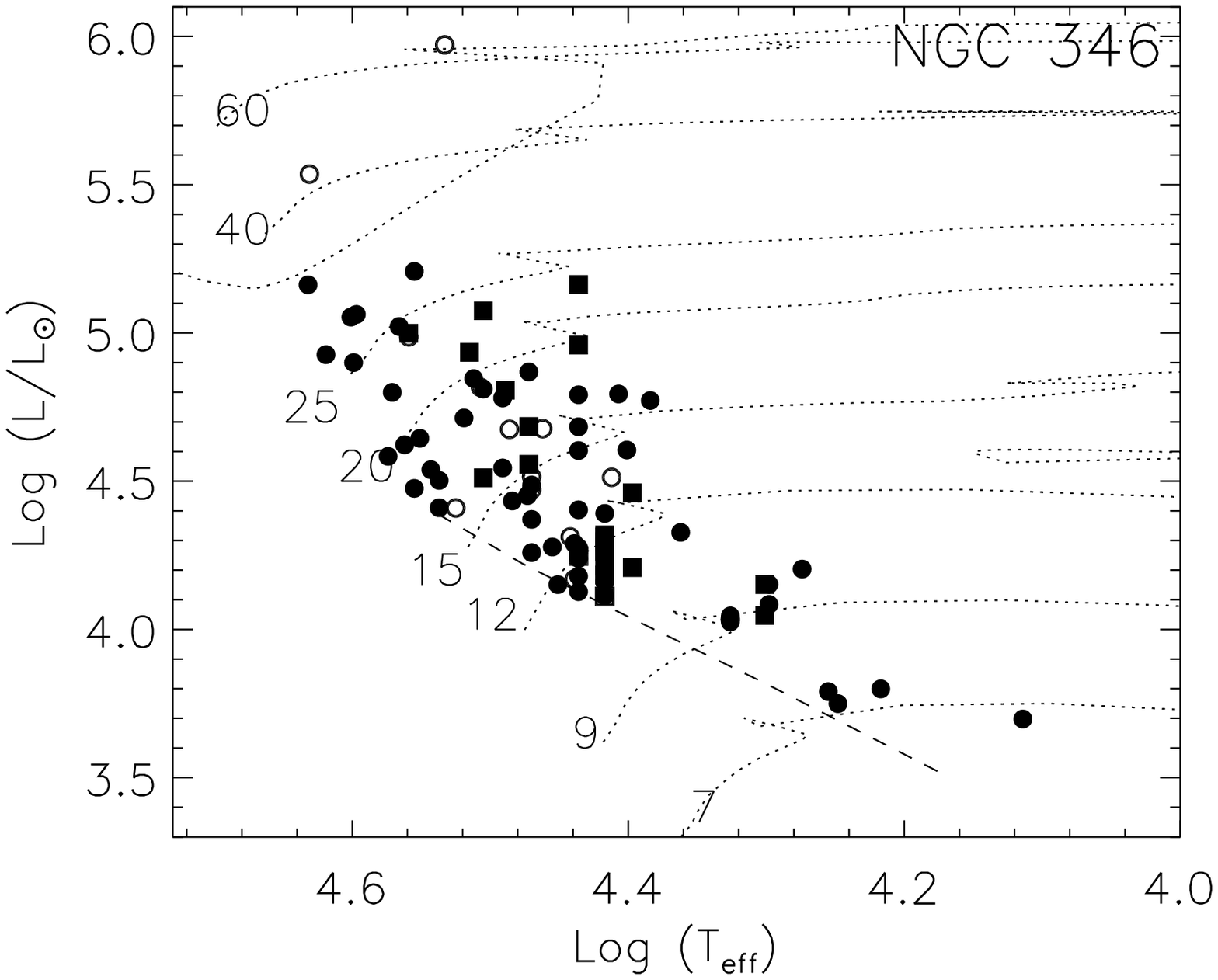, height=70mm, angle=0}\\
\epsfig{file=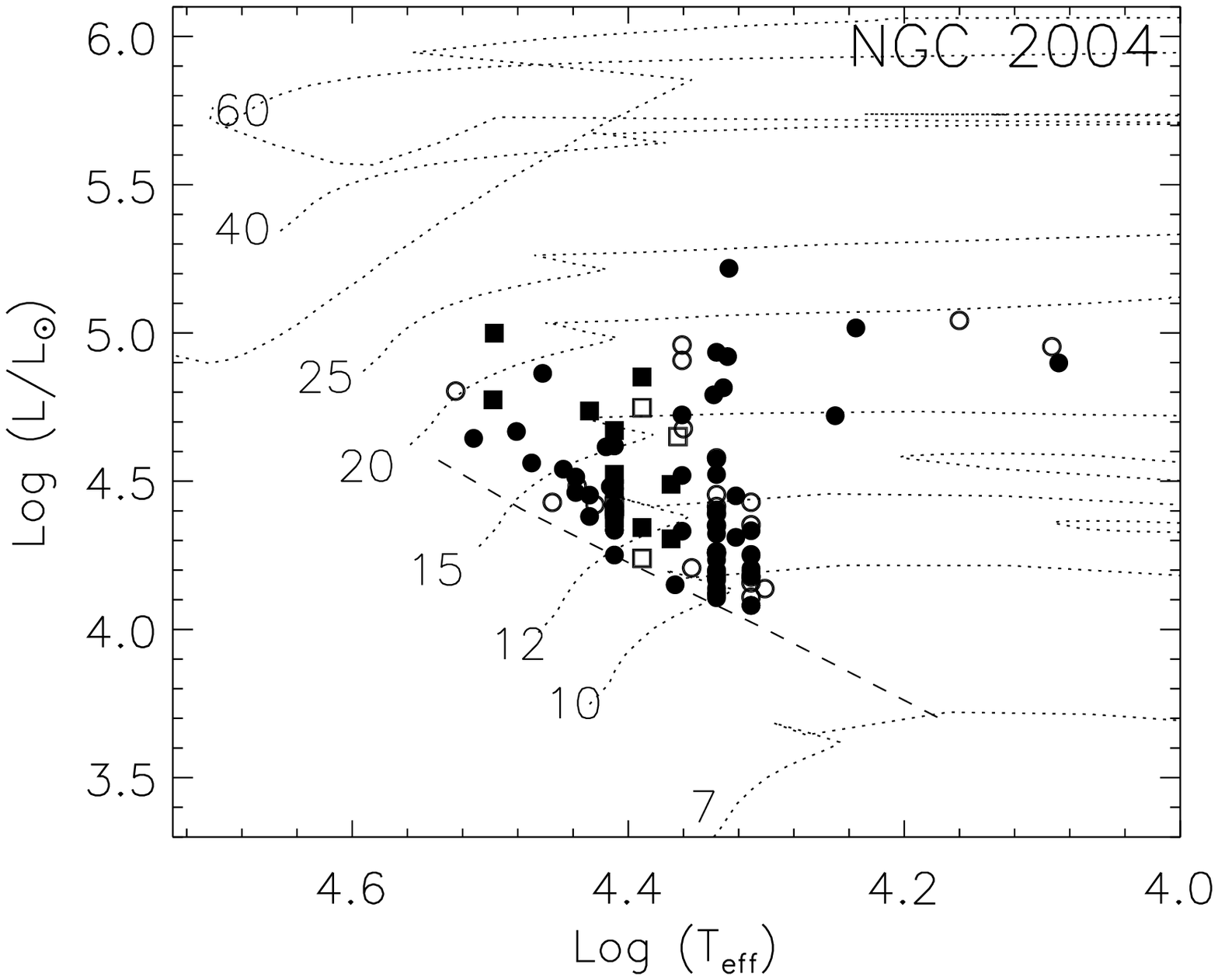, height=70mm, angle=0}&
\epsfig{file=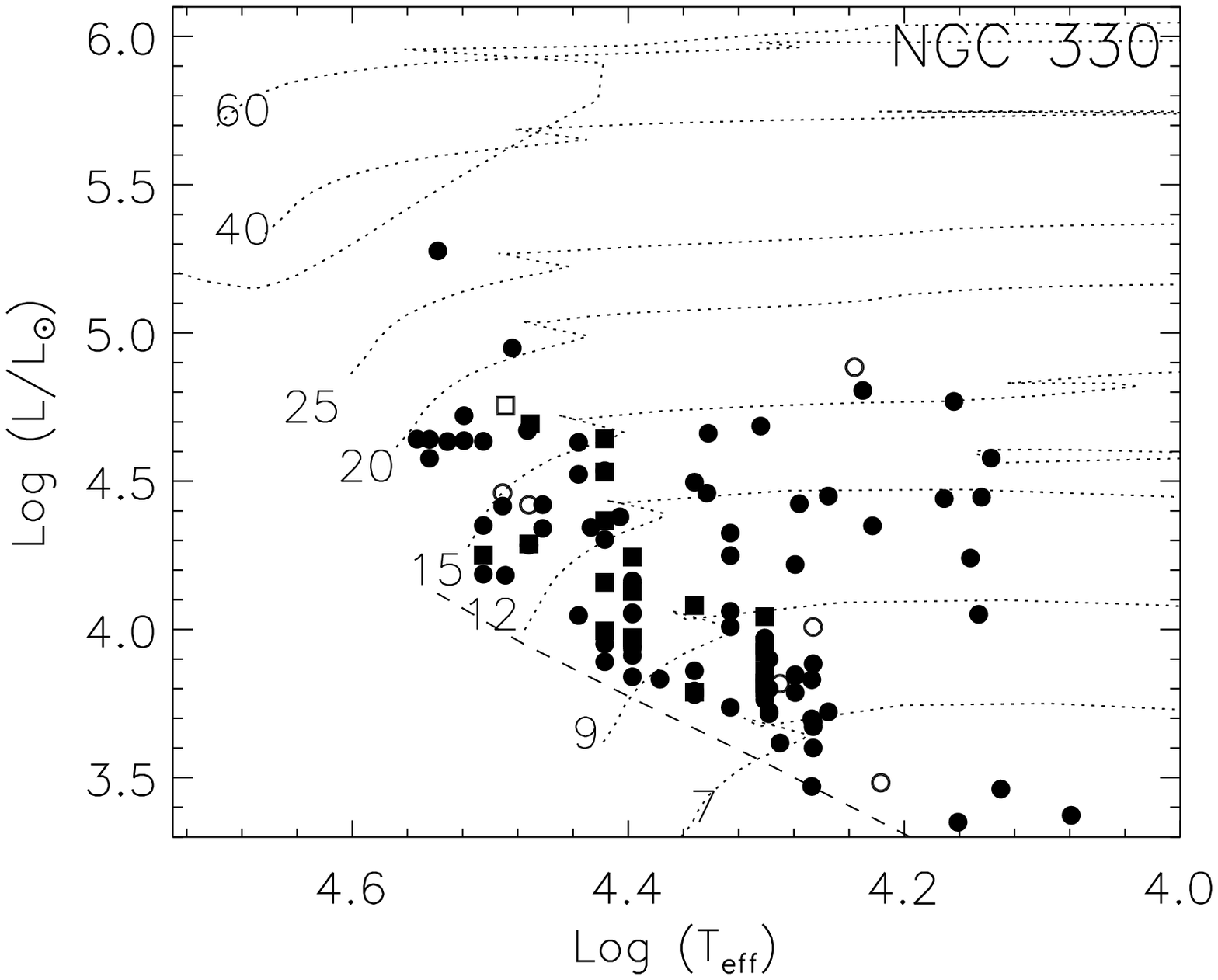, height=70mm, angle=0}\\
\end{tabular}
\caption[]{Hertzsprung-Russell diagrams for our sample of objects towards N\,11,
NGC\,2004, NGC\,346 and NGC\,330. Be objects are plotted as
squares and objects showing evidence of binarity are plotted as open symbols.
Non-rotating evolutionary tracks have been obtained from Meynet et al. 
(\cite{mey94}) and
Schaerer et al. (\cite{sch93}) for the LMC objects and from Meynet et al. and
Charbonnel et al. (\cite{cha93}) for the SMC objects. The dashed line indicates
the magnitude limit for our observations in each region. 
It should be noted that
many of the objects in each sample lie in overlapping positions in these
figures.}
\label{f_HR}
\end{figure*}

\subsection{Masses}

In Fig.~\ref{f_HR}, Hertzsprung-Russell diagrams are presented for the sample of
stars towards each cluster. Non-rotating evolutionary tracks are also shown
and from these the evolutionary masses of each object have been 
estimated and these are listed 
in Tables~\ref{t_346_vsini}-\ref{t_2004_vsini}. Isochrones are not included
as our sample is likely to be dominated by field stars
and therefore the populations are unlikely to be coeval, see
Sect.~\ref{s_clustermem}.

The dashed line in Fig.~\ref{f_HR} indicates the magnitude limit for our
observations in each region (see Paper II)
and it is clear that we sample the zero-age main-sequence (ZAMS) 
at differing masses in
each clusters. For example we tend to sample more evolved
objects in NGC\,2004 due to our target slection, but 
do sample stars on the ZAMS below 9, 12 and 15 M$_{\rm \sun}$ 
in NGC\,330, NGC\,346 and N\,11 respectively. 

The uncertainty in the masses for those objects for which we have estimated
effective temperatures
is typically 10-15\%, which arises {\bf principally}
from the uncertainty in the luminosity. In
Tables~\ref{t_346_vsini}-\ref{t_2004_vsini} we do not give atmospheric
parameters for Be stars, double lined spectroscopic binaries and stars with
uncertain spectral types. However, for the estimation 
of their masses, we
assume temperatures based on their spectral type (or their mean spectral type
where a spectral type range is given). The uncertainty in the mass of
these objects is typically 30\%. In addition to these evolutionary masses it is
possible to estimate spectroscopic masses, but an uncertainty of 0.25\,dex
in the 
surface gravity estimate corresponds to an uncertainty of over 70\% in the
spectroscopic mass and hence such estimates are of limited value.

\subsubsection{Mass discrepancy}  \label{s_mass_discrep}

Although individual spectroscopic masses are highly uncertain, 
our sample is large enough to examine systematic differences between
evolutionary and spectroscopic masses. For example, it is well known that the
evolutionary masses derived for
OB-type supergiants typically are greater 
than the spectroscopic masses (see for
example, Herrero et al., \cite{her92}; Trundle \& Lennon, \cite{tru05}; 
Lennon et al., \cite{len03}). 
In Table~\ref{t_massdis} the average mass discrepancies are
summarized. While the discrepancy for evolved
stars has previously been observed, that
for the main-sequence objects is unexpected. 
Although the scatter is large, given the
number of objects, there may be a systematic difference. If it is assumed that
no mass discrepancy should exist for main-sequence objects since this
evolutionary stage is thought to be well understood, that observed 
could be due to binarity, errors in the distance estimate, bolometric
corrections, reddening values or even the derived surface gravities. However,
given that the magnitude of the effect is different in the SMC compared to the
LMC, it is unlikely to be due to the derived gravities, or bolometric
corrections as the same methodologies were used for each. 
Additionally the mass discrepancies for the binary systems are 
almost identical to those in Table~\ref{t_massdis} indicating that undetected
binary systems are unlikely to be the cause of the discrepancies.
Since the reddening
values are small, the discrepancy may come from the distance estimates.

\begin{table}[ht]
\caption[Mass discrepancy]{Mean mass-discrepancies for the sample of {\bf single B-type} stars.
{\bf Objects are classified by surface gravity with supergiants
having gravities less than 3.2\,dex (see Sect.~\ref{s_evolve}), giants 
between 3.2 and 3.7\,dex  and dwarfs greater than 3.7\,dex.}
Errors are the standard deviations and the value in brackets indicates
the number of objects.}
\label{t_massdis}
\centering
\begin{tabular}{lcc}\\ \hline \hline
                 & \multicolumn{2}{c}{log (M$_{\rm evolutionary}$/M$_{\rm
spectroscopic})$}\\
		     & LMC & SMC\\
\hline
\\
Supergiants      & ~0.17$\pm$0.17 (22) & ~0.25$\pm$0.19 (13) \\
Giants           & ~0.06$\pm$0.14 (43) & ~0.18$\pm$0.14 (39) \\
Dwarfs           & -0.14$\pm$0.22 (51) & -0.05$\pm$0.17 (60) \\
\\
\hline
\end{tabular}
\end{table}

Using the evolutionary masses and atmospheric parameters listed in
Tables~\ref{t_346_vsini}-\ref{t_2004_vsini}, we have derived luminosity
estimates. It should be noted that these are different from the luminosities
listed in Tables~\ref{t_346_vsini}-\ref{t_2004_vsini}, which assumed a fixed
distance to each Cloud.
Adopting the same
reddenings and bolometric corrections as discussed in Sect.~\ref{s_lum},
it is then possible to derive distance estimates 
towards each star. For objects with
$\log g \geq$\,3.7\,dex the mean distance moduli to the LMC and SMC are
18.16$\pm$0.77 and 18.79$\pm$0.55 from 51 and 60 objects respectively, where
the uncertainties are the 1$\sigma$ standard deviations, compared with our
adopted values of 18.56 and 18.91. Although
the mean value for the LMC is 0.4\,dex smaller than that adopted in
Sect.~\ref{s_lum},
it lies within the range
of those reported for the LMC in the recent literature.
For example,
Udalski (\cite{uda98}) estimates a distance modulus of 18.13 for the LMC.
However, it should also be noted that Udalski similarly estimates a distance
modulus of 18.63\,dex for the SMC which is 0.17\,dex closer than our derived value of
18.80\,dex {\bf and 0.28\,dex closer than our adopted value of 18.91\,dex. Excluding 
the most massive
stars (greater than 25\,$_{\rm \sun}$) increases the mass discrepancies 
although within the uncertainties this is not significant.}
To maintain consistency with published
work from the FLAMES project the 
distance estimates discussed in Sect.~\ref{s_lum} have been adopted
throughout {\bf the subsequent discussion.}

If corrections are applied to remove the mass discrepancy 
of the dwarf stars it further enhances the mass discrepancy of the evolved
objects. However, such corrections result in 
similar mass discrepancies for the
LMC and SMC evolved samples.
This may indicate that the post-main-sequence evolution of massive
stars is not fully understood and as discussed above, such mass discrepancies
have previously been
observed. In Sect.~\ref{s_evolve} 
several
inconsistencies between the observations and the theoretical predictions of
evolved objects are highlighted. 
In particular, we argue for
extension of the hydrogen burning phase of the evolutionary tracks
which would result in lower evolutionary
masses being derived, hence reducing the mass discrepancy for evolved stars.


\section{{\bf Populations within the sample}}     \label{s_nature}

{\bf Atmospheric parameters and rotational velocities 
have been derived for a large
sample of objects which contains several groups with
different properties. In
this section we compare these various subsets using as a control sample
single, core hydrogen burning stars of less
than 25\,M$_{\rm \sun}$. In Sect.~\ref{s_discussion}
we discuss how the subsets can be utlised
to answer the three questions listed in Sect.~\ref{s_intro}.}

\subsection{High mass stars}                                  \label{s_vmass}

Mass-loss rates generally increase with stellar mass (Maeder \& Meynet
\cite{mae01}) and
hence more massive objects should lose more angular momentum and rotate slower
than less massive objects for a given initial rotational velocity.
The theoretical evolutionary tracks of Maeder \& Meynet
show that mass-loss
effects become significant for masses greater than 25\,M$_{\rm \sun}$.
In Table~\ref{t_mass} we compare the
mean projected rotational velocity for stars with masses greater than
25\,M$_{\rm \sun}$ and those with masses of $\leq$\,25\,M$_{\rm \sun}$.
There is indeed some evidence that the more massive stars rotate slower,
although
it should be stressed that the number of objects with masses greater than
25\,M$_{\rm \sun}$ is {\bf only 7\% of the sample} and 
the trend may not be statistically significant.

\begin{table}
\centering
\caption[]{Mean projected rotational velocities {\bf for single core-hydrogen
buring stars} with masses greater
than 25\,M$_{\rm \sun}$ and masses of $\leq$\,25\,M$_{\rm \sun}$ 
in each region. Values in brackets indicate the number of stars in each
sample.}
\label{t_mass}
\begin{tabular}{lcccc}
\hline \hline
Region    & \multicolumn{4}{c}{Mean $v\sin i$ (km\,s$^{-1}$)}         \\
          &\multicolumn{2}{c}{M\,$\leq$\,25\,M$_{\rm \sun}$} 
                              &\multicolumn{2}{c}{M\,$>$\,25\,M$_{\rm \sun}$}\\
\hline \\
NGC\,346  &  150                 & (65)& 113                   & ( 5)\\
NGC\,330  &  154                 & (77)& ~73                   & ( 1)\\
N\,11     &  149                 & (34)& 126                   & (12)\\
NGC\,2004 &  121                 & (60)& ~--                   & ( 0)\\
\\ \hline
\end{tabular}
\end{table}

\subsection{Binarity}                                      \label{s_binarity}

Binarity effects can significantly alter the evolution of massive stars. For
example, mass-transfer can lead to spin-up of the accreting star, while tidal
interaction in close binary systems can lead to reduced
rotational velocities through partial or full synchronisation of the rotational
velocity with the orbital velocity (see for example 
Huang \& Gies \cite{hua06}; Abt et al. \cite{abt02}; Zahn \cite{zah77}). 
Indeed both Zahn (\cite{zah94}) 
and Hunter et al. (\cite{hun06}) find that massive
stars in binary systems typically have smaller amounts of core processed
material in their photosphere compared to single stars. This is 
consistent with
tidal interaction reducing the rotational velocities and hence suppressing
the rotational mixing of the products of nucleosynthesis into the photosphere.

\begin{table}[ht]
\centering
\caption[]{{\bf Mean projected rotational velocities of the 
core hydrogen burning
single and binary
stars less massive than 25\,M$_{\rm \sun}$ in each region. The binary
population includes both single and double lined spectroscopic
binaries.}
Values in brackets indicate the number of stars in each
sample. }
\label{t_binary}
\begin{tabular}{lcccc}
\hline \hline
Region    & \multicolumn{4}{c}{Mean $v\sin i$ (km\,s$^{-1}$)}         \\
 & \multicolumn{2}{c}{Single stars} & \multicolumn{2}{c}{Binary stars}\\
\hline \\
NGC\,346  &  147                 & (67)& 103                   & (25)\\
NGC\,330  &  146                 & (83)& 121                   & ( 8)\\
N\,11     &  143                 & (35)& ~92                  & (30)\\
NGC\,2004 &  120                 & (62)& ~99                  & (28)\\
\\ \hline
\end{tabular}
\end{table}

{\bf In Table~\ref{t_binary} the mean projected rotational velocities 
of single and
binary stars are compared. It can be seen that binaries tend to have lower
rotational velocities than single stars. This may indicate that tidal forces
do play an important part in reducing the rotational velocities of massive
stars.}
However, it is important to consider selection effects. Radial velocity
variations (the evidence of binarity) are detected by cross-correlating
individual exposures of the same wavelength region observed at different epochs
to look for radial velocity shifts (Sect.~\ref{s_rad_corr}). Such shifts are
relatively simple to detect to a few km\,s$^{-1}$ for stars with projected
rotational velocities up to 100\,km\,s$^{-1}$ as the absorption lines are
typically sharp and well defined. For stars with broad absorption lines  and
hence large projected
rotational velocities, it is difficult to define the position of the core
of the line and small radial velocity shifts will remain undetected.
Additionally in binary systems with small orbital periods, 
the separation between
the objects would be smaller, the velocity shifts
greater and tidal interaction would be more important
than for longer period systems. As such
the rotational velocity distributions of binary objects could 
be biased
towards stars having low rotational velocities. Clearly further studies of the
binary populations in young clusters are required.

%

\subsection{Supergiants} \label{s_supergiants}

{\bf In Table~\ref{t_evolve} the mean projected rotational
velocities of the supergiant and core hydrogen buring objects
are compared with supergiants being considered as objects
with surface gravities less than 3.2\,dex (see Sect.~\ref{s_evolve}). 
Assuming that the inclination axes of our
targets are randomly orientated the mean rotational velocities would 
be a factor
of 4/$\pi$ larger than the values quoted in Table~\ref{t_evolve}.
It is clear that supergiants have much lower rotational velocities than
core hydrogen burning objects which is simply a consequence of their
evolutionary status.}

\begin{table}
\centering
\caption[]{Mean projected rotational velocities of the single 
supergiant {\bf (surface gravities of less than 3.2\,dex) and core
hydrogen burning stars} with masses $\leq$25\,M$_{\rm \sun}$.
Values in brackets indicate the number of stars 
in each sample. Note that as upper limits are included in the
supergiant sample, the means 
given here should be treated as upper limits to this class of object.}
\label{t_evolve}
\begin{tabular}{lcccc}
\hline \hline
Region    & \multicolumn{4}{c}{Mean $v\sin i$ (km\,s$^{-1}$)}         \\
          & \multicolumn{2}{c}{Supergiants} & \multicolumn{2}{c}{Core hydrogen}\\
          &       &     & \multicolumn{2}{c}{burning stars}\\
\hline \\
NGC\,346  &  72   & ( 1)& 147		& (67)  \\
NGC\,330  &  38   & (12)& 146  		& (83)  \\
N\,11     &  60   & (11)& 143  		& (35)  \\
NGC\,2004 &  60   & ( 8)& 119  		& (62)  \\
\\ \hline
\end{tabular}
\end{table}

\subsection{Be stars} \label{s_Be}

It has recently been reported that Be stars typically have larger rotational
velocities than B-type stars (Martayan et al. \cite{mar06a}, \cite{mar06b}).
A significant number (17\%) of the
stars in the FLAMES survey have been classified as Be type objects and hence it
is
possible to compare the rotational velocities of normal B-type stars with those
showing Be characteristics. In Table~\ref{t_Be} we present the average projected
rotational velocities of the B- and Be-type stars in each region.

\begin{table}
\centering
\caption[]{{\bf Comparison of the mean projected rotational velocities 
of the single core hydrogen burning B- and Be-type
stars with masses $\leq$25\,M$_{\rm \sun}$ in each region. }
Values in brackets indicate the number of stars in each
sample. }
\label{t_Be}
\begin{tabular}{lcccc}
\hline \hline
Region    & \multicolumn{4}{c}{Mean $v\sin i$ (km\,s$^{-1}$)}         \\
          & \multicolumn{2}{c}{B stars} & \multicolumn{2}{c}{Be stars}\\
\hline \\
NGC\,346  &  115                 & (45)& 213                   & (22)\\
NGC\,330  &  128                 & (64)& 207                   & (19)\\
N\,11     &  134                 & (31)& 215                   & ( 4)\\
NGC\,2004 &  105                 & (51)& 190                   & (11)\\
\\ \hline
\end{tabular}
\end{table}

It is apparent that on average Be-type stars have higher rotational velocities than
normal B-type stars in agreement with the finding of Martayan et al.
(\cite{mar06a}, \cite{mar06b}), {\bf see also Fig.~\ref{f_marcomp}}. 
Zorec et al. (\cite{zor05}) have shown that
the Be phenomena normally
occurs early in the core hydrogen burning phase. 
They also show that for the most massive 
stars the Be phenomena disappears during their main-sequence lifetimes
and
postulate that this is due to mass-loss causing a star to
spin down.

Table~\ref{t_Be} additionally shows that we have observed many more Be-type stars in
the SMC than the LMC, again in agreement with Martayan et al. \cite{mar06b}. 
If B-type stars become Be-type stars because 
they reach the main-sequence
with sufficiently high angular momentum, this may indicate that at lower
metallicity, massive stars tend to rotate faster and suffer a smaller amount of
spin down. 


\section{Discussion}                                        \label{s_discussion}

\subsection{Constraints on evolutionary history}     \label{s_evolve}

\begin{figure*}[ht]
\centering
\begin{tabular}{cc}
\epsfig{file=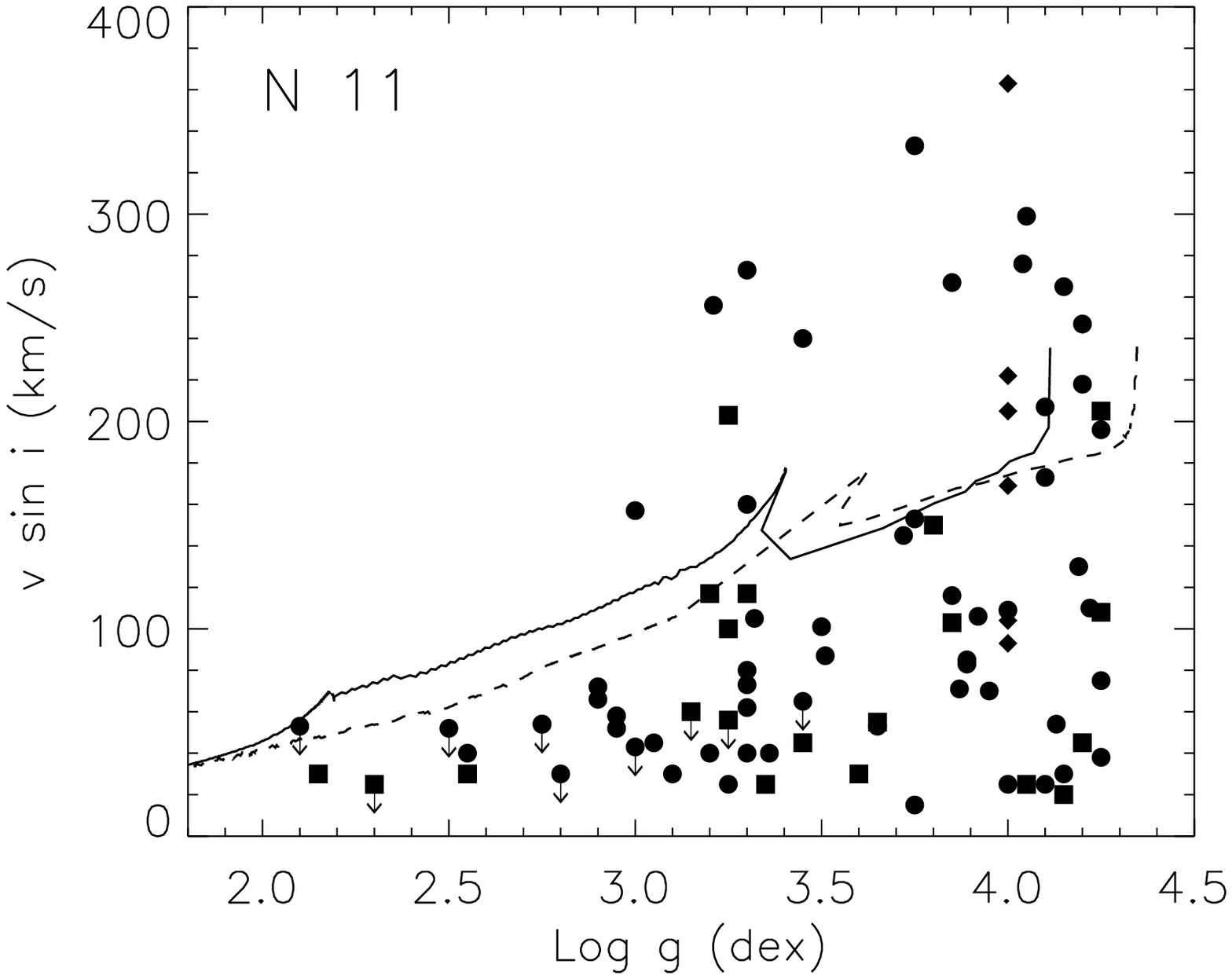, height=60mm, angle=0} &
\epsfig{file=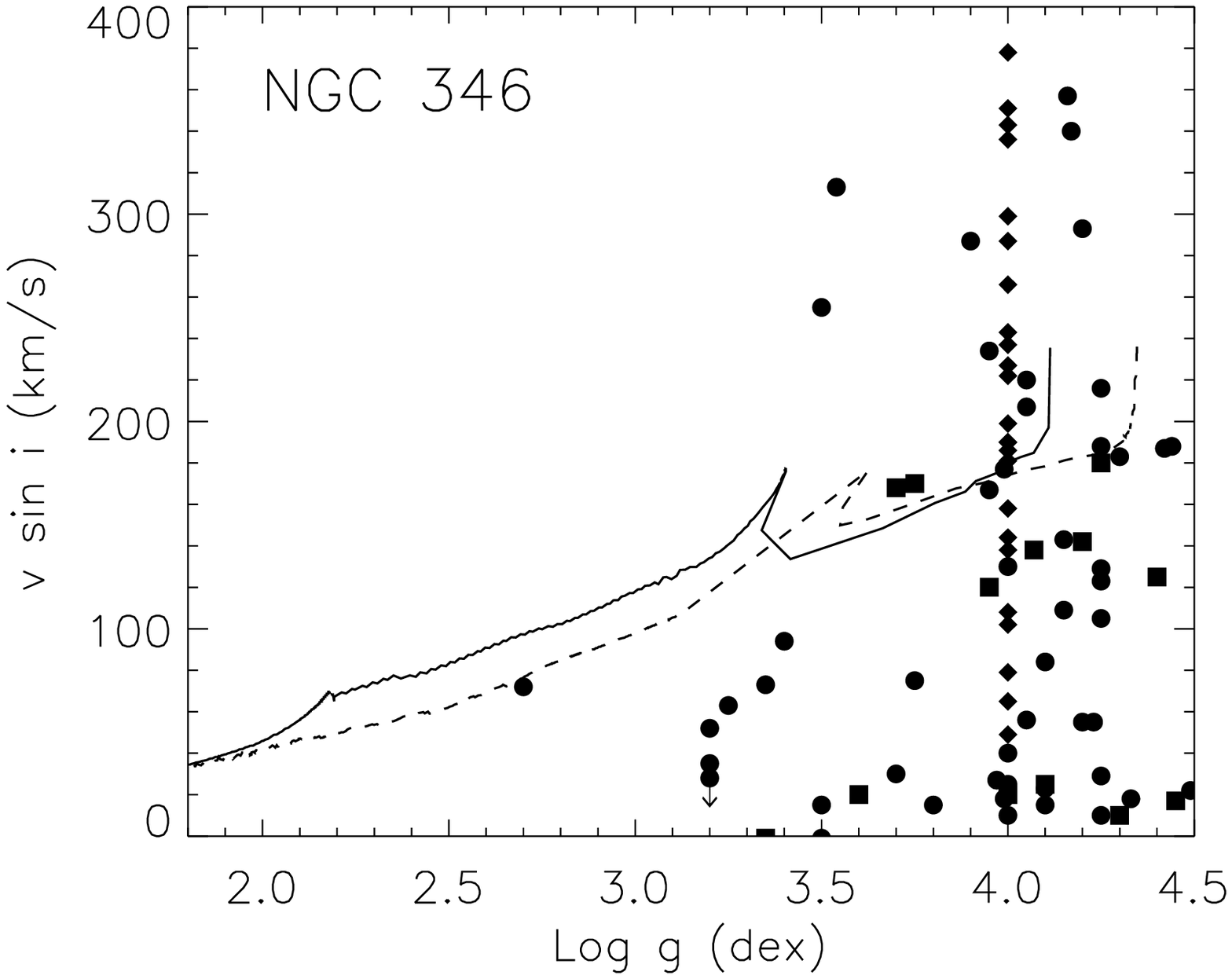, height=60mm, angle=0} \\
\epsfig{file=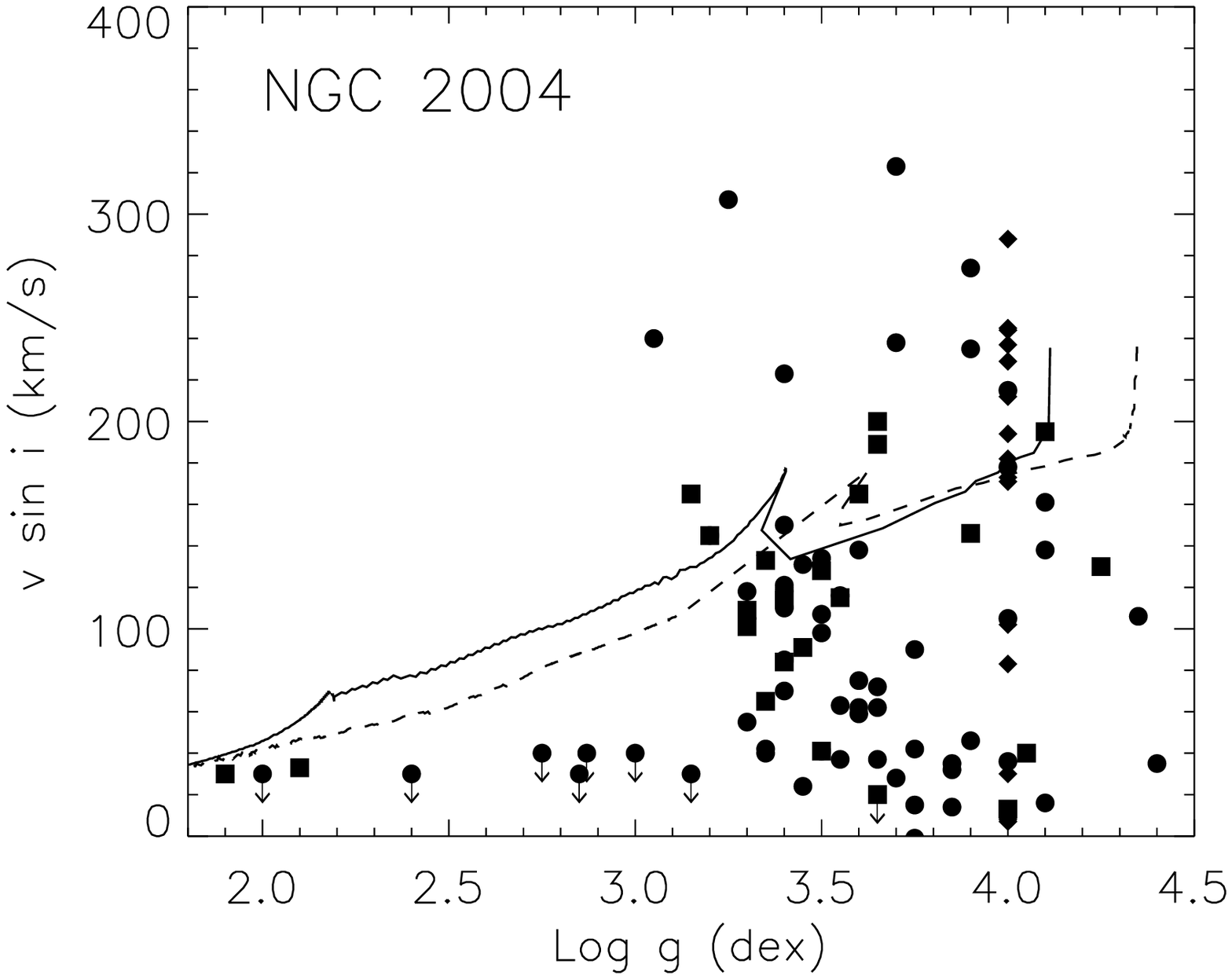, height=60mm, angle=0}&
\epsfig{file=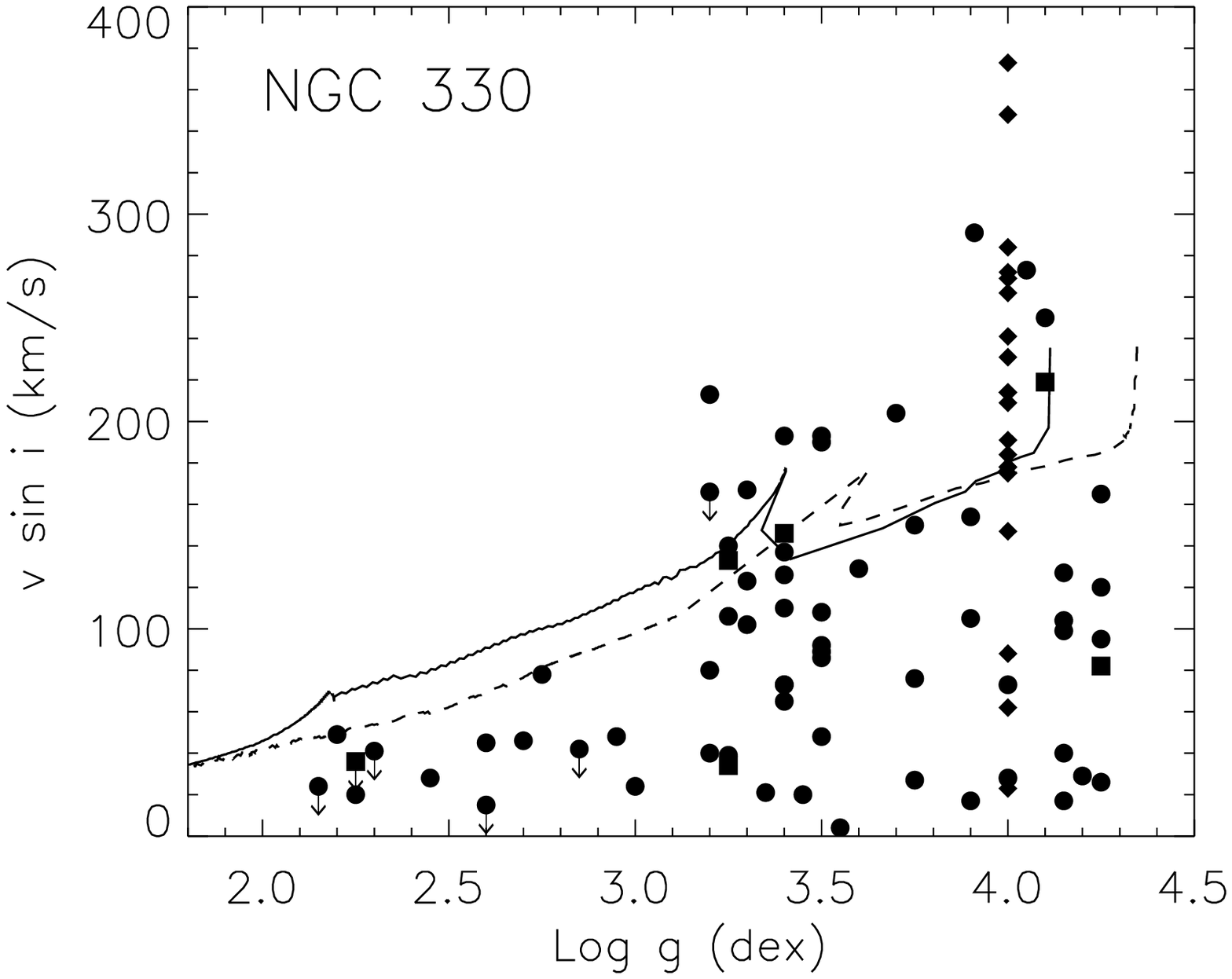, height=60mm, angle=0}\\
\end{tabular}
\caption[]{Plots of projected rotational velocity ($v\sin i$) 
against the surface gravity 
for stars towards N\,11, NGC\,2004, NGC\,346 and
NGC\,330 {\bf showing the bi-modal distribution of
rotational velocities}. Be objects are assumed to be main-sequence objects with a surface 
gravity of 4.00\,dex and are plotted as diamonds. Single lined
spectroscopic binaries are indicated
by square symbols. Downward arrows 
indicate upper limits to the $v\sin i$. The
stellar evolutionary tracks of Maeder \& Meynet (\cite{mae01})
for a 15\,M$_{\rm \sun}$ star, with an initial rotational velocity of
300\,km\,s$^{-1}$ (scaled to take into account random angles of inclination)
at SMC (dashed line) and Galactic (solid line) metallicity are shown. }
\label{f_logg_vsini}
\end{figure*}

The surface gravity of an object is a good indicator of its evolutionary status,
with the
gravity decreasing as an object evolves from a dwarf, to a giant and then into a
supergiant star. 
In Fig.~\ref{f_logg_vsini} we plot the projected
rotational velocity against surface gravity. 

The distributions of rotational velocities shown in Fig.~\ref{f_logg_vsini}
appear
bi-modal in the sense that for surface gravities of $<$3.2\,dex 
the projected rotational velocities
are small, while at higher values the distribution
is much broader.  {\bf We interpret this limit of 3.2\,dex to be the
end of the core hydrogen burning phase},
since after this point the stars rapidly expand to become
slowly rotating, core helium burning objects, as represented
by the late B-type objects in our sample with surface gravities of
$\sim$2.0\,dex. {\bf Hence we consider objects with surface gravities lower than
3.2\,dex to be supergiants}.
We note that the predicted
position of the end of core hydrogen burning shown by the evolutionary
tracks occurs at higher surface gravities than
that implied by our observations and suggests that the predicted 
core hydrogen burning 
lifetime needs to be extended to lower surface gravities. It is
however important to stress a caveat. If the mass-discrepancy
discussed in Sect.~\ref{s_mass_discrep}  is not due to an error in the adopted
distance moduli, but
instead due to a systematic error in our estimated surface gravities then
our targets would 
have surface gravities that are systematically $\sim$0.1\,dex higher. 

A large population of the giants lie in the post-main-sequence 
gap when
compared with the evolutionary tracks shown in Fig.~\ref{f_HR}.
Due to their small evolutionary timescale, such objects would not normally be
expected to be observed.
A similar over abundance of
post-main-sequence objects in the Galactic clusters NGC\,3293 and NGC\,4755
was observed in Paper III. 
Extension of the end of the main-sequence (for example, by increasing
the efficiency of overshooting) would allow the
evolutionary tracks to encompass this giant phase during the
core hydrogen burning phase. 
This would then be consistent {\bf with the} presence
of such objects in our samples.

The theoretical predictions of
Maeder \& Meynet (\cite{mae01}) indicate that rotational velocities
remain relatively constant during core hydrogen burning, decreasing by
approximately 40\,km\,s$^{-1}$ from the zero-age main-sequence to the end of
hydrogen burning. From Fig.~\ref{f_logg_vsini} there may be 
some evidence that
the objects with the largest surface gravities cover the widest range of
rotational velocities {\bf and indeed objects classified as giants have rotational
velocities that are lower than those of dwarfs by on} average
$\sim$50\,km\,s$^{-1}$ for our entire sample. This appears to be consistent with
giants being in the core hydrogen burning phase, again implying that extension
of the evolutionary tracks may be necessary.

The evolutionary tracks of Maeder \& Meynet (\cite{mae01}) which incorporate
rotation show that for a 15\,M$_{\rm \sun}$ star it takes $\sim$13.6\,Myr
to reach the end of the hydrogen burning main-sequence, while the blue
supergiant phase has a lifetime of less than 0.1\,Myr, {\bf i.e. the blue supergiant
phase is less than 1\% of the core hydrogen burning lifetime. The large number
of supergiants observed in the FLAMES survey is unexpected with the ratio of supergiants to 
core hydrogen burning objects being $\sim$11\%.}
Significant
populations of objects in the post-main-sequence gap
have previously been seen (see, for
example, Caloi et al. \cite{cal93}; Fitzpatrick \& Garmany \cite{fit90}). However,
it should be noted that the FLAMES
survey preferentially selected the brightest objects towards each cluster, and
hence there may be some bias towards selecting supergiants.
\begin{figure*}[ht]
\centering
\begin{tabular}{c}
\epsfig{file=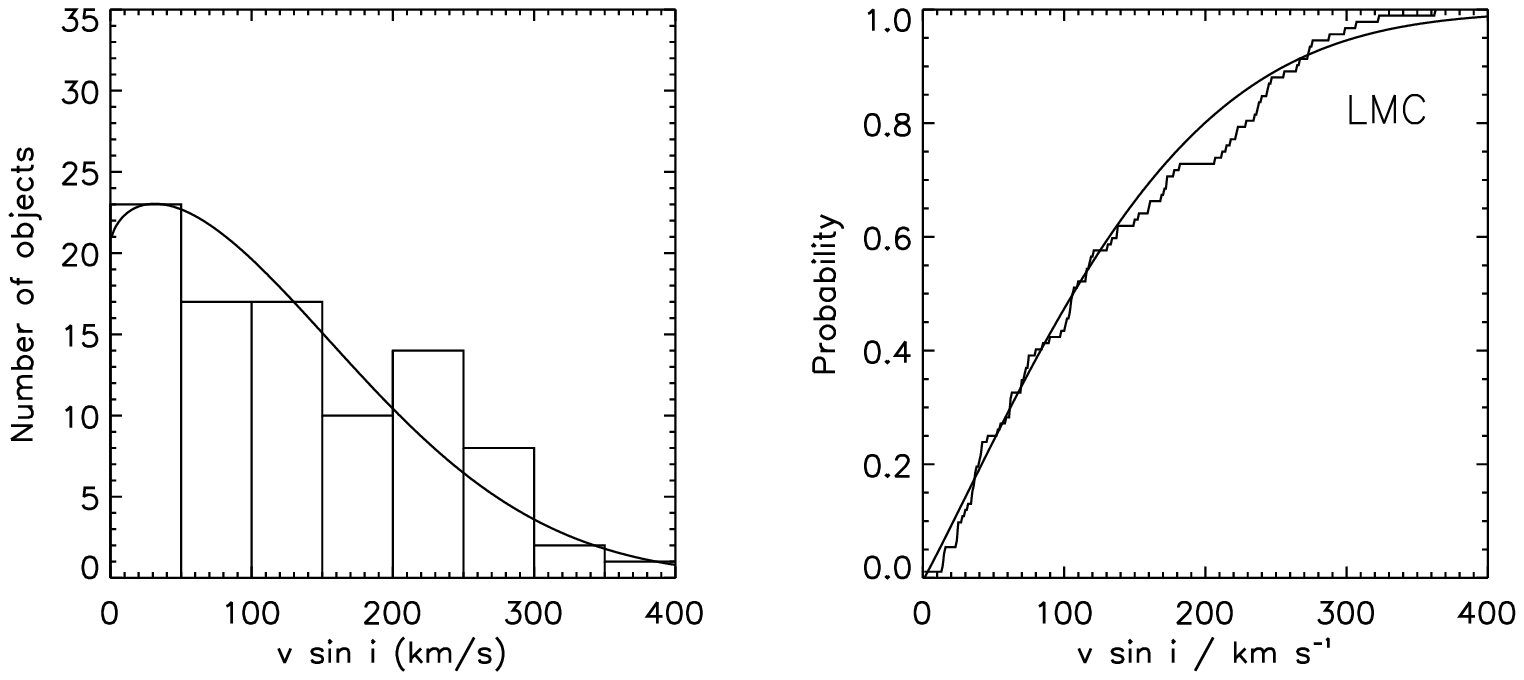, height=70mm, angle=0}\\
\epsfig{file=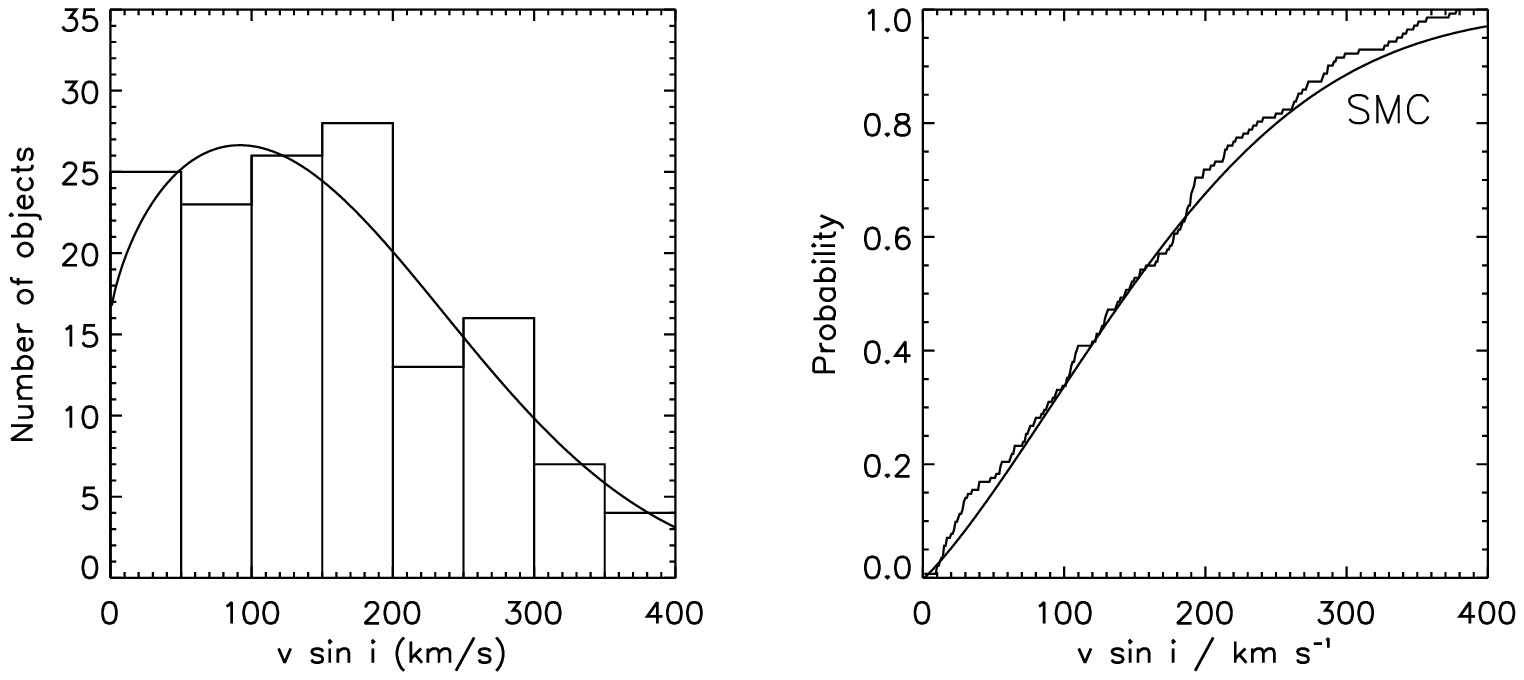, height=70mm, angle=0}\\
\end{tabular}
\caption[]{Rotational velocity distributions of {\bf the single core-hydrogen burning
LMC and SMC stars with masses $\leq$25\,M$_{\rm \sun}$, upper
and lower panels respectively.} The LMC distribution is fitted 
with a intrinsic
rotational velocity modelled by a Gaussian with a peak at 100\,km\,s$^{-1}$ and
a $\frac{1}{e}$ half width, 
of 150\,km\,s$^{-1}$, while the SMC distribution is
fitted with a Gaussian having a peak at 175\,km\,s$^{-1}$ and a half-width of
150\,km\,s$^{-1}$.}
\label{f_hist_gauss}
\end{figure*}

The evolutionary tracks shown in Fig.~\ref{f_logg_vsini} imply that the
rotational velocity slows down relatively smoothly 
 as the star evolves to low
surface gravities after the core hydrogen burning phase, 
although, {\bf as discussed above}, this phase is almost
instantaneous compared with the core hydrogen burning lifetime. 
However, observationally this
trend is not seen; objects with surface gravities $<$3.2\,dex 
have relatively constant rotational velocities, which are smaller than predicted.
A similar trend has
previously been observed by Dufton et al. (\cite{duf06b}). 
However, as discussed by Hunter
et al. (\cite{hun06}) and Trundle et al. (\cite{tru07}) many of the supergiant
objects are highly enriched in nitrogen 
and show evidence of binarity, suggesting that a
mass-transfer event may have occurred. As such, comparison with single star
evolutionary tracks may not be 
valid. Indeed, rejuvenation through a mass-transfer
event is one method of populating this short evolutionary stage and may
reconcile the over-abundance of supergiants in our samples with the predicted
lifetime of such an evolutionary stage from single star models. 

An alternative
explanation is the possibility 
of blue-loops, where the supergiant goes through a
red-supergiant phase and then returns to higher effective temperatures. 
Such a process
would lead to the observed nitrogen enhancements and rotational velocities 
and an increased lifetime in
the blue supergiant phase although the blue-loop phase is not well constrained
in the evolutionary models (G. Meynet and N. Langer, private communications).
We therefore believe that the supergiant population of objects shown in
Fig.~\ref{f_logg_vsini} are not the result of direct single star
evolution from the {\bf core hydrogen burning phase and either binarity
or blue loops may be responsible for this supergiant population}.


%

\subsection{Rotation and metallicity} \label{s_vmet}

As discussed in Sect.~\ref{s_intro}, stars at low metallicity are predicted
to rotate faster as they lose less angular
momentum through their stellar winds and are generally more compact.
Our sample of objects covers the metallicity regimes of the LMC and SMC and
hence it is possible to look for metallicity effects. To increase
our sample sizes we have combined together those for N\,11
and NGC\,2004 in the LMC and NGC\,346 and NGC\,330 in
the SMC sample. {\bf Only single core hydrogen burning objects with masses
of $\leq$25\,M$_{\rm \sun}$ have been considered and this corresponds to
48\% and 70\% of our LMC and SMC samples
respectively. The upper limit in mass was set to minimise the possible effects
due to mass-loss as discussed in Sect.~\ref{s_vmass}, whilst the supergiants
have been excluded as their rotational velocities will not be representative of
their main-sequence progenitors. Our samples generally have 
spectral types between B0 and B3, 
and have initial masses between 
7 and 25\,M$_{\rm \sun}$. The spectral type distributions
of both samples are similar and there is no reason to expect
any trend of rotational velocity with spectral type over this small range. 
}

In Fig.~\ref{f_hist_gauss}, the distribution of projected
rotational velocities in the LMC and SMC are shown together with their
cumulative probability
functions. The
number of objects we have
observed should be sufficient to allow us to estimate
the intrinsic rotational velocity distribution.
Following the procedures discussed in Paper III, we have attempted to fit
these distributions with a Gaussian function convolved with a function to take
into account random angles of inclination. 
A chi-squared test has been performed
to obtain the best fit to the cumulative probability functions. 
We chose not to apply the chi-squared test to the fit to the 
histograms, as this may be dependent on the choice of binsize.

{\bf For the LMC sample,} an intrinsic velocity distribution 
with a peak at 100\,km\,s$^{-1}$ and 
a $\frac{1}{e}$ half width of 150\,km\,s$^{-1}$ (which corresponds
to a to a full-width-half-maximum, FWHM, of 250\,km\,s$^{-1}$) {\bf 
is found, with corresponding values for the SMC sample being
175\,km\,s$^{-1}$ and 150\,km\,s$^{-1}$.} Although 
a Gaussian distribution has been assumed, as
discussed in Paper III and in Mokiem et al. (\cite{mok06})
any function with a similar profile would also provide
a satisfactory  fit. For example as the Gaussian profiles are quite broad, 
a rectangular profile centred on 
the Gaussian peaks, with similar widths, would also give a reasonable
fit to the observed distributions.
From these results it appears that the LMC objects
typically rotate significantly more slowly than those of the SMC. Indeed,
the Student's t-test gives a 
93\% probability that the means of the two distributions
are significantly different while the K-S statistical test gives the 
probability that
the distributions are drawn from the same parent population to be less than 10\%.

To further investigate
metallicity effects the Magellanic Cloud samples can be compared with
Galactic results. However before this is
possible, it is necessary to ensure that 
the samples are comparable. For example, the Galactic objects observed in
the FLAMES project
and discussed in Paper III mostly lie within two cluster radii of their
respective clusters. From the discussion in Sect.~\ref{s_clustermem} 
it is apparent that the objects that we
observed in the Magellanic Clouds lie much further from the centre of their
respective associations. Our sample is therefore likely to be dominated
by field or unbound objects and it may be more
appropriate to compare the rotational velocity distributions discussed here with
Galactic field stars, rather than the Galactic cluster stars discussed in Paper
III. Indeed, it has been shown in Paper III and by Wolff et al. (\cite{wol07}) 
that the mean rotational velocities
of Galactic cluster objects are higher than Galactic field stars. {\bf For example,
strong peaks in the Galactic cluster rotational velocity distributions occur at 
$\sim$200\,km\,s$^{-1}$ while the field population discussed below peaks 
at $\sim$50\,km\,s$^{-1}$.}

\begin{figure}[ht]
\centering
\epsfig{file=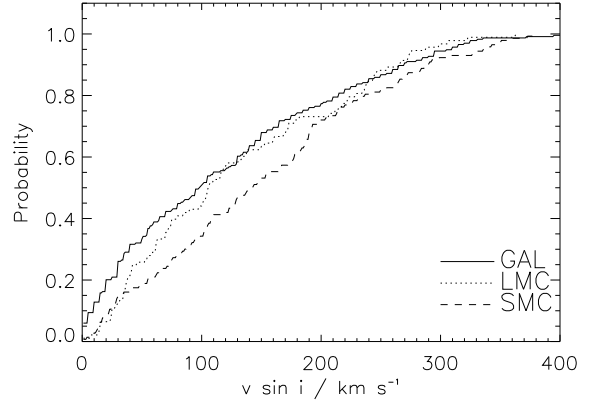, height=60mm, angle=0}\\
\caption[]{Cumulative probability functions of the rotational velocities of
the {\bf single core-hydrogen burning LMC and SMC objects with masses 
$\leq$25\,M$_{\rm \sun}$.} For 
comparison Galactic (GAL) 
objects from Howarth et al. (\cite{how97}) and Abt et al. 
(\cite{abt02}) are included (see text). The median intrisic rotational
velocities are 125, 135 and 183\,km\,s$^{-1}$ 
for the Galactic, LMC and SMC samples respectively (assuming
random angles of inclination) {\bf indicating a metallicity dependence}.}
\label{f_metallicity_cdf}
\end{figure}

Abt et al. (\cite{abt02}) and Howarth et al. (\cite{how97}) 
present rotational velocities for a large number
of B-type and O-type stars in the Galactic field respectively.
In order to maintain consistency between 
the Galactic and Magellanic Cloud samples we have randomly chosen Galactic
objects from these catalogues in such a way as to populate an equivalent
spectral type distribution as the Magellanic Cloud samples, with 198
objects being selected in total.
In Fig.~\ref{f_metallicity_cdf} cumulative probability functions are plotted for
this Galactic sample and the 
LMC and SMC objects shown in Fig.~\ref{f_hist_gauss}. It is clear that the SMC
objects typically rotate faster than the Galactic objects with the LMC
distribution lying between the two. 
Given that we are randomly selecting Galactic
comparison objects, we have carried out the comparison 1000 times. The Student's
t-test and the K-S test give the probability that the Galactic and LMC samples 
are drawn from the same distribution  to be
43\% and 24\% respectively from the mean of these 1000
comparisons. Similarly the two tests show that the probability the
Galactic and SMC samples are drawn from the same distribution of rotational
velocities to be 0.3\% and 0.8\%. Hence while there is some evidence that the
LMC sample of stars may
rotate faster that those in the Galactic field, 
the faster rotation of the SMC B-type
objects compared to the Galactic objects is significant at the 3$\sigma$ level.

\subsection{Implications for gamma-ray burst progenitors}                       \label{s_GRBs}

The progenitors of long duration gamma-ray bursts (hereafter GRBs)
have been suggested to be rapidly rotating massive carbon-oxygen
stars that have lost their entire hydrogen and helium envelopes. This
poses a problem for stellar evolutionary models in that one has to
remove the outer envelope of a massive star, to create a Wolf-Rayet
type object (WR), 
but maintain or enhance the rotational velocity (Woosley \& Heger
\cite{wooheg06}). 

Very rapidly rotating main-sequence stars have been proposed as candidates
for chemically homogeneous evolution, in which the stars are homogeneously
mixed on a timescale much shorter than their nuclear lifetimes
(Maeder \cite{maeder87}; Langer \cite{langer92}).
In  this case the star becomes a WR star directly by evolving to the blue, 
and its outer hydrogen envelope is effectively mixed into the star
rather than being lost through stellar winds. This avoids 
angular momentum loss through mass-loss, and hence one can potentially
get a viable GRB progenitor i.e. a massive CO star that is hydrogen and 
helium free and rapidly rotating (Yoon \& Langer \cite{yoo05}, Woosley \& Heger
\cite{wooheg06}). Our rotational velocity results, and the determination
of a frequency distribution can provide a critical input parameter for
these models. 

Yoon et al. (\cite{yoo06}) have recently generated massive star evolutionary 
models at SMC and lower metallicities to study the plausibility of 
chemically homogeneous evolution generating GRB 
progenitors. They assumed a distribution function of rotational velocities 
which was estimated from the most massive O-type stars in our FLAMES survey 
data set. Mokiem et al. (\cite{mok06}) determined projected rotational
velocities 
for the 17 unevolved O-stars 
in NGC\,346 and estimated the intrinsic rotational velocity 
distributions, as in Sect.~\ref{s_vmet}. 
Given the young age of NGC\,346 it would be expected that angular momentum loss
due to winds should be small and the current rotational velocity distribution
should be representative of the initial rotational velocity distribution. 
In order to determine the rate of GRB production from rapidly rotating, 
chemically homogeneous stars, Yoon et al. (\cite{yoo06}) 
then assumed that this rotational velocity 
distribution was {\bf appropriate to} all masses and metallicities. They 
predict that at metallicities of $Z=0.00001-0.001$ stars with initial masses
as low as 11-14\,M$_{\rm \sun}$ can produce GRBs. By combining our full
FLAMES  survey sample (Mokiem et al., and this paper) 
we have the most extensive rotational 
velocity survey at low metallicities to test these ideas. 

\begin{figure}[ht]
\centering
\epsfig{file=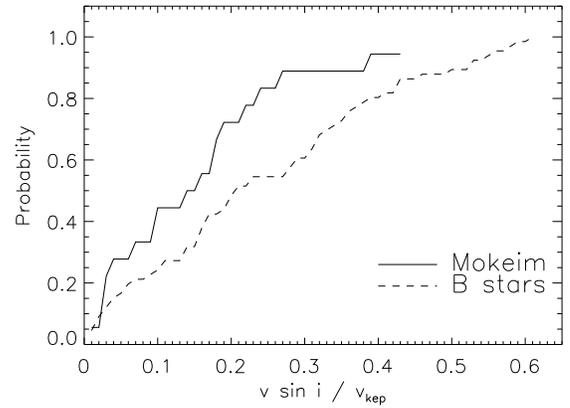, height=60mm, angle=0}\\
\caption[]{Cumulative probability functions of projected rotational velocity to
critical rotational velocity ratio for the unevolved NGC\,346 O-star
sample from Mokiem et al. (\cite{mok06}, Mokiem) 
and our {\bf single core-hydrogen burning} NGC\,346 B-type star sample. {\bf This plot
shows that the rotational velocity distribution of the more
massive O-stars is peaked towards lower rotational velocities than that of the
B-type stars}.}
\label{f_grb}
\end{figure}

As suggested by Table~\ref{t_mass} the more massive stars 
may rotate slower than their less massive
counterparts, implying a possible mass dependence. However, if mass-loss
effects are small in the O-star sample whilst on the
main-sequence, this effect
may be due to mass and metallicity dependent protostellar winds
rather than winds on the main-sequence (Yoon et al. \cite{yoo06}).
In Fig.~\ref{f_grb}
we plot the cumulative probability function for the ratio of 
projected rotational
velocity to critical (Keplerian) rotational velocity for the unevolved
O-stars in the Mokiem et al. NGC\,346 sample and compare this to that derived
for the NGC\,346 B-type stars here. There is a clear offset between the two
distributions, 
with the B-stars having higher ratios of rotational to critical velocity. 
We note again that the mass range of the B-stars presented here are
10-25\,M$_{\rm \sun}$, whereas the O-stars of Mokiem et al. have 
masses mostly in the range 20-60\,M$_{\rm \sun}$. 
{\bf This is the first time we have had enough objects to carry out
this mass-dependent comparison. 
Hence the Yoon et al. rates of GRBs from rapidly rotating stars below 
20\,M$_{\rm \sun}$ (greater than $v \sin i / v_{\rm Kep} \simeq 0.5$)
are probably underestimates. }

{\bf It would be desirable to recalculate GRB
rates with our new measurements of lower mass
stellar rotational velocities.}  Yoon et al. (in their Fig.~3) calculate that
stars of 30\,M$_{\rm \sun}$ require a minimum rotational velocity of
$v sin i/v_{\rm kep} \sim 0.27$ to be viable GRB candidates, which
from Fig.~\ref{f_grb} is 
about 10-15\% of the O-type stellar population.  A 14\,M$_{\rm
\sun}$ star however requires a minimum rotational velocity of $v sin
i/v_{\rm kep} \sim 0.43$, suggesting about 15\% of that population
could be viable GRB progenitors.  Given the similarity of these
percentages, the initial mass function should imply more progenitors from
the lower mass range.  
{\bf We therefore suggest that adopting our observed rotational velocity for 
the full mass range will increase the fraction of stars which
become GRBs. A quantatative estimate 
requires a detailed grid calculation as in Yoon et
al. This is outside the scope of this paper but it should be 
done with the complete rotational velocity distribution that we present
here}.

\section{Conclusions} \label{s_conclude}

Atmospheric parameters and projected 
rotational velocities are presented for approximately
400 O- and early B-type stars towards four clusters in the Magellanic Clouds,
which combines this work, the analysis of some 50 O-type stars (Mokiem et al.
\cite{mok06}, \cite{mok07}) and the analysis of 100 narrow lined B-type stars
(Hunter et al. \cite{hun06} and Trundle et al. \cite{tru07}). 
This represents the full data set from our ESO Large Programme, the 
VLT Flames survey of massive stars in the Magellanic clouds and the Galaxy. 
{\bf From the subsets of the data we find that supergiants are the slowest 
rotators in the sample, typically
having rotational velocities less than 80\,km\,s$^{-1}$. Additionally binaries
tend to rotate slower than single
objects, but given that binarity is easier to detect at low projected
rotational velocities this result may be dominated by selection effects.
It is also
found that objects with Be phenomena are typically the fastest rotators in our
dataset, which is in agreement with previous studies. 
The number of high mass objects (M$>$25\,M$_{\rm \sun}$) in
the combined sample is limited but there is some evidence that the
most massive objects rotate slower than their less massive
counterparts. We utilise the single
core-hydrogen burning objects along with these subsets to address the
questions outlined in Sect.~\ref{s_intro}}.

\begin{itemize}
\item

{\bf Current evolutionary models do not predict the large
number of giants observed in our samples which may imply that the theoretical
tracks need to be extended. Indeed we find that
stars classed as giants
typically rotate some $\sim$50\,km\,s$^{-1}$ slower than dwarf objects and this
is consistent with a modest slowing down of the rotational velocity over the
core hydrogen burning lifetime, 
indicating that giants are still in the hydrogen burning
phase. From the bi-model distribution of rotational velocities for supergiant
and core-hydrogen burning objects we observe the end of the core-hydrogen 
burning phase to be at 3.2\,dex. We also show that supergiants are unlikely to
have evolved directly from the hydrogen burning main-sequence, but instead, binarity 
or blue loops may be a more likely explanation for the observed supergiant 
population.
}

\item
Our sample is dominated by field objects and we compare the rotational velocity
distributions with Galactic field objects from the literature, finding that the
SMC metallicity stars rotate fastest, in agreement with theoretical
predictions. The difference between the SMC stars and Galactic
field stars is significant at the 3$\sigma$ level, suggesting we have detected
a real difference. 
However we find no significant difference between the
rotational velocity distributions of the Galactic and LMC stars. 
 By assuming random angles of inclination, we can model the observed 
$v \sin i$ distributions with broad Gaussians with the parameters listed 
in  Table~\ref{t_mcvels}.

\begin{table}
\centering
\caption[]{Peak and width of the Magellanic Cloud rotational velocity
distributions determined by fitting the distributions with a Gaussian function,
see Sect.~\ref{s_vmet}.}
\label{t_mcvels}
\begin{tabular}{lcc} \hline \hline
Region & Peak & $\frac{1}{e}$ half-width \\
       & km\,s$^{-1}$ & km\,s$^{-1}$ \\
\hline \\
LMC    & 100  & 150 \\
SMC    & 175  & 150 \\
\\ \hline
\end{tabular}
\end{table}

\item 
{\bf Objects more massive than 25\,M$_{\rm \sun}$ tend to rotate
slower than less massive objects and 
this} could be due to mass-loss effects, with a very
massive star losing angular momentum through strong stellar winds,
although this is unexpected in the SMC. An alternative explanation is
that the initial birth rotational velocity is mass dependent, perhaps
due to protostellar winds.  {\bf However, the consequence
is that Yoon et al. (\cite{yoo06}) may have underestimated 
the rotational velocity distribution of 10-20\,M$_{\rm \sun}$
GRB progenitors.} We therefore suggest 
that the rotational velocity distribution derived here should be
employed in future GRB rate calculations, and this may even increase the rate
of GRBs expected at low metallicty due to {\bf the greater prevalence of
B-type stars}. 

\end{itemize}

\begin{acknowledgements}
We are grateful to staff from the European Southern Observatory 
for assistance in
obtaining the data. This work, conducted as part of the award `Understanding 
the lives of
massive stars from birth to supernovae' (SJS) made under the
European Heads of Research Councils and European Science Foundation
EURYI (European Young Investigator) Awards scheme, was supported by
funds from the Participating Organisations of EURYI and the EC Sixth
Framework Programme. SJS also acknowledges the Leverhulme Trust.  Additionally
we acknowledge support from
PPARC (UK Particle Physics and Astronomy Research
Council) and DEL (Dept. 
of Employment \& Learning
Northern Ireland). 

\end{acknowledgements}


\newpage

\begin{table*}
\addtocounter{table}{-13}
\centering
\caption{Radial distances ($r_{\rm rad}$, arcmins)
of the objects in the NGC\,346 FLAMES sample with
respect to the centre of star formation taken as the position of object 435 
from the Massey et al. (\cite{mas89}) catalogue of OB stars in NGC\,346;
$\alpha$\,=\,00\,59\,04.49 $\delta$\,=\,-72\,10\,24.7. The
centre of NGC\,346 adopted in Paper II for calculation of radial distances
was obtained from
the SIMBAD database (operated at the CDS, Strasbourg, France) and is actually 
the centre of the ionised shell of gas, which is offset from the cluster
centre.}
\label{t_346_rad}
\begin{tabular}{cccccccc}
\hline \hline
Object & $r_{\rm rad}$ & Object & $r_{\rm rad}$ & 
 Object & $r_{\rm rad}$ & Object & $r_{\rm rad}$ \\
\hline \\
NGC\,346-001&2.13&NGC\,346-030&4.27&NGC\,346-059&5.53&NGC\,346-088&6.69\\
NGC\,346-002&5.55&NGC\,346-031&7.01&NGC\,346-060&4.75&NGC\,346-089&8.07\\
NGC\,346-003&3.36&NGC\,346-032&5.83&NGC\,346-061&5.68&NGC\,346-090&5.09\\
NGC\,346-004&7.22&NGC\,346-033&0.71&NGC\,346-062&6.61&NGC\,346-091&8.68\\
NGC\,346-005&8.49&NGC\,346-034&0.44&NGC\,346-063&5.69&NGC\,346-092&4.14\\
NGC\,346-006&5.69&NGC\,346-035&5.84&NGC\,346-064&5.44&NGC\,346-093&0.74\\
NGC\,346-007&0.56&NGC\,346-036&2.15&NGC\,346-065&1.88&NGC\,346-094&5.28\\
NGC\,346-008&5.74&NGC\,346-037&4.07&NGC\,346-066&1.86&NGC\,346-095&8.43\\
NGC\,346-009&3.74&NGC\,346-038&5.34&NGC\,346-067&6.19&NGC\,346-096&8.38\\
NGC\,346-010&6.88&NGC\,346-039&8.60&NGC\,346-068&5.60&NGC\,346-097&0.86\\
NGC\,346-011&9.17&NGC\,346-040&3.46&NGC\,346-069&9.43&NGC\,346-098&2.70\\
NGC\,346-012&4.82&NGC\,346-041&8.42&NGC\,346-070&3.56&NGC\,346-099&8.63\\
NGC\,346-013&2.34&NGC\,346-042&8.39&NGC\,346-071&7.85&NGC\,346-100&6.89\\
NGC\,346-014&4.99&NGC\,346-043&4.02&NGC\,346-072&3.32&NGC\,346-101&4.96\\
NGC\,346-015&8.47&NGC\,346-044&3.25&NGC\,346-073&7.21&NGC\,346-102&5.34\\
NGC\,346-016&6.04&NGC\,346-045&3.89&NGC\,346-074&8.30&NGC\,346-103&7.54\\
NGC\,346-017&7.54&NGC\,346-046&3.80&NGC\,346-075&4.68&NGC\,346-104&1.19\\
NGC\,346-018&2.93&NGC\,346-047&9.35&NGC\,346-076&1.88&NGC\,346-105&3.50\\
NGC\,346-019&2.37&NGC\,346-048&2.55&NGC\,346-077&1.31&NGC\,346-106&7.49\\
NGC\,346-020&6.41&NGC\,346-049&7.70&NGC\,346-078&5.43&NGC\,346-107&0.45\\
NGC\,346-021&4.58&NGC\,346-050&1.48&NGC\,346-079&0.34&NGC\,346-108&4.59\\
NGC\,346-022&1.32&NGC\,346-051&0.37&NGC\,346-080&1.14&NGC\,346-109&6.03\\
NGC\,346-023&1.94&NGC\,346-052&6.93&NGC\,346-081&5.61&NGC\,346-110&5.91\\
NGC\,346-024&2.67&NGC\,346-053&7.60&NGC\,346-082&5.78&NGC\,346-111&0.49\\
NGC\,346-025&3.73&NGC\,346-054&2.88&NGC\,346-083&1.83&NGC\,346-112&1.76\\
NGC\,346-026&3.87&NGC\,346-055&3.20&NGC\,346-084&5.88&NGC\,346-113&4.10\\
NGC\,346-027&3.12&NGC\,346-056&1.06&NGC\,346-085&6.63&NGC\,346-114&5.16\\
NGC\,346-028&2.56&NGC\,346-057&5.38&NGC\,346-086&0.20&NGC\,346-115&0.11\\
NGC\,346-029&1.76&NGC\,346-058&3.40&NGC\,346-087&6.19&NGC\,346-116&6.26\\
\\ \hline
\end{tabular}
\end{table*}

\begin{table*}
\addtocounter{table}{1}
\caption{Atmospheric parameters and $v\sin i$ values for the NGC\,346 sample. 
In the method column, M06, T06 or H06 indicates that the values have
been 
directly 
obtained from Mokiem et al. (\cite{mok06}, \cite{mok07}), 
Trundle et al. (\cite{tru07})
or Hunter et al. (\cite{hun06}). He
indicates the methodology outlined in Sect.~\ref{s_parameters} has been used
to derived the atmospheric parameters. `A' indicates that the temperature
has been assumed based on spectral type. Values marked with colons indicate
parameters where fitting has been difficult and the uncertainties may be larger
than normal. Where it was possible to
measure the projected rotational velocity of the secondary object in a double
lined spectroscopic binary system, a second entry is given with `B' appended to
the object identifier. We do not give atmospheric parameters for those objects
listed as Be stars (Be), double lined spectroscopic binaries (SB2)
 or classified with
an uncertain  spectral type (SP?).} 
\label{t_346_vsini}
\centering
\begin{tabular}{llcccccccl}\hline \hline
Star&Spectral Type&T$_{\rm eff}$&$\log g$&$v\sin i$&Method&L/L$_{\rm \sun}$&Mass
                                                                    &Comments \\
             &              & (K)&(dex) &(km\,s$^{-1})$&&  &(M$_{\rm \sun}$)&        \\
\hline \\
NGC\,346-001 & O7 Iaf+      & 34100& 3.35 &  74      & M06& 6.02 & 66 & Binary\\
NGC\,346-004 & Be(B1:)      &   Be & --   & 266      & -- & 5.17 & 23 &       \\
NGC\,346-007 & O4 V((f+))   & 42800& 3.95 & 120      & M06& 5.45 & 40 & Binary\\
NGC\,346-008 & B1e          &   Be & --   & 299      & -- & 4.96 & 20 &       \\
NGC\,346-009 & B0e          &   Be & --   & 199      & -- & 5.08 & 24 &       \\
NGC\,346-010 & O7 IIIn((f)) & 35900& 3.54 & 313      & M06& 5.20 & 27 &       \\
NGC\,346-012 & B1 Ib        & 24200& 3.20 & $<$28    & H06& 4.77 & 16 &       \\
NGC\,346-013 & B1:          &  SB2 &  --  & 120      & -- & 4.79 & 17 & Binary\\
NGC\,346-013B&              &  SB2 &  --  & 320      & -- &	 &    & Binary\\
NGC\,346-015 & B1 V         &  SB2 &  --  &  45      & -- & 4.79 & 17 & Binary\\
NGC\,346-015B&              &  SB2 &  --  &  35      & -- &	 &    & Binary\\
NGC\,346-016 & B0.5 Vn      &  SB2 &  --  & 181      & -- & 4.87 & 19 & Binary\\
NGC\,346-018 & O9.5 IIIe    & 32700& 3.33 & 138      & M06& 5.10 & 24 &       \\
NGC\,346-020 & B1 V+earlyB  &  SB2 &  --  &  34      & -- & 4.69 & 16 & Binary\\
NGC\,346-020B&              &  SB2 &  --  &  66      & -- &	 &    & Binary\\
NGC\,346-021 & B1 III       & 25150& 3.50 &  15      & H06& 4.61 & 14 &       \\
NGC\,346-022 & O9 V         & 36800& 4.20 &  55      & M06& 4.95 & 24 &       \\
NGC\,346-023 & B0.2:(Be-Fe) &   Be &   -- &  65      & -- & 4.81 & 18 &       \\
NGC\,346-024 & B2:shell(Be-Fe)& Be &   -- & 190      & -- & 4.59 & 19 &       \\
NGC\,346-025 & O9V          & 36200& 4.07 & 138      & M06& 4.90 & 23 & Binary\\
NGC\,346-026 & B0 IV (Nstr) & 32500& 3.75 &  75      & He & 4.85 & 20 &       \\
NGC\,346-027 & B0.5 V       & 31000& 4.05 & 220      & He & 4.78 & 18 &       \\
NGC\,346-028 & OC6 Vz       & 42900& 3.97 &  27      & M06& 5.10 & 32 &       \\
NGC\,346-029 & B0 V         & 32150& 4.10 &  25      & H06& 4.82 & 19 & Binary\\
NGC\,346-030 & B0 V         &  SB2 &  --  & 183      & -- & 4.81 & 19 & Binary\\
NGC\,346-031 & O8 Vz        & 39500& 3.99 &  18      & M06& 4.99 & 27 &       \\
NGC\,346-032 & B0.5 V       & 29000& 4.40 & 125      & He & 4.68 & 17 & Binary\\
NGC\,346-033 & O8 V         & 39900& 4.44 & 188      & M06& 4.99 & 27 &       \\
NGC\,346-035 & B1 V         &  SB2 &   -- & 145      & -- & 4.61 & 15 & Binary\\
NGC\,346-035B&              &  SB2 &   -- & 105      & -- &	 &    & Binary\\
NGC\,346-036 & B0.5 V(Be-Fe) & Be &   -- & 287      & -- & 4.69 & 17 &       \\
NGC\,346-037 & B3 III       & 18800& 3.20 &  35      & H06& 4.21 & 10 &       \\
NGC\,346-039 & B0.7 V       & 25800& 3.60 &  20      & H06& 4.51 & 13 & Binary\\
NGC\,346-040 & B0.2 V       & 30600& 4.00 &  20      & H06& 4.67 & 17 & Binary\\
NGC\,346-041 & B2 (Be-Fe)  &   Be &   -- & 144      & -- & 4.46 & 12 &       \\
NGC\,346-043 & B0 V         & 33000& 4.25 &  10      & H06& 4.71 & 18 &       \\
NGC\,346-044 & B1 II        & 23000& 3.50 &  40      & H06& 4.33 & 11 &	      \\
NGC\,346-045 & B0.5 Vne     &   Be &   -- & 181      & -- & 4.56 & 15 &	      \\
NGC\,346-046 & O7 Vn        & 39700& 4.17 & 340      & M06& 4.81 & 24 &	      \\
NGC\,346-047 & B2.5 III     & 19850& 3.25 &  63      & A  & 4.15 & 10 &	      \\
NGC\,346-048 & Be (B3 shell)&   Be &   -- & 158      & -- & 4.15 & 10 &	      \\
NGC\,346-049 & B8 II        & 13000& 2.70 &  80      & He & 3.70 &  7 &	      \\
NGC\,346-050 & O8 Vn        & 37200& 4.16 & 357      & M06& 4.67 & 21 &	      \\
NGC\,346-051 & O7 Vz        & 41600& 4.33 &  18      & M06& 4.87 & 27 &	      \\
NGC\,346-052 & B1.5 V       &  SB2?&   -- &  12      & -- & 4.37 & 12 &Binary?\\
NGC\,346-053 & B0.5 V       & 29500& 3.75 & 170      & He & 4.51 & 15 & Binary\\
NGC\,346-054 & B1 V         & 27300& 4.20 &  23      & A  & 4.41 & 13 &	      \\
NGC\,346-055 & B0.5 V       & 29500& 4.00 & 130      & He & 4.49 & 15 &	      \\
NGC\,346-056 & B0 V         & 31000& 3.80 &  15      & H06& 4.55 & 16 &	      \\
\\
\hline
\end{tabular}
\end{table*}

\begin{table*}
\addtocounter{table}{-1}
\caption{-continued.}
\centering
\begin{tabular}{llcccccccl}\hline \hline
Star&Spectral Type&T$_{\rm eff}$&$\log g$&$v\sin i$&Method&L/L$_{\rm \sun}$&Mass
                                                                    &Comments \\
             &              & (K)&(dex) &(km\,s$^{-1})$&&  &(M$_{\rm \sun}$)&        \\
\hline \\
NGC\,346-057 & B2.5 III     & 19850& 3.35 &  73      & A  & 4.09 &  9 &	      \\
NGC\,346-058 & B0.5 V       & 29500& 4.25 & 180      & He & 4.47 & 14 & Binary\\
NGC\,346-061 & B1-2 (Be-Fe) &   Be &   -- & 336      & -- & 4.32 & 12 &	      \\
NGC\,346-062 & B0.2 V       & 29750& 4.00 &  25      & H06& 4.45 & 15 &	      \\
NGC\,346-064 & B1-2 (Be-Fe) &   Be &   -- & 108      & -- & 4.31 & 12 &	      \\
NGC\,346-065 & B3 (Be-Fe)   &   Be &   -- & 222      & -- & 4.05 &  9 &	      \\
NGC\,346-066 & O9.5 V       & 35600& 4.25 & 129      & M06& 4.59 & 19 &	      \\
NGC\,346-067 & B1-2 (Be-Fe) &   Be &   -- & 351      & -- & 4.29 & 12 &	      \\
NGC\,346-068 & B0 V (Be-Fe) &   Be &   -- & 378      & -- & 4.51 & 16 &	      \\
NGC\,346-069 & B1-2 (Be-Fe) &   Be &   -- & 186      & -- & 4.28 & 12 &	      \\
NGC\,346-070 & B0.5 V       & 30500& 4.15 & 109      & He & 4.43 & 15 &	      \\
NGC\,346-072 & B1-2 (Be-Fe) &   Be &   -- & 102      & -- & 4.26 & 12 &	      \\
NGC\,346-073 & B1-2 (Be-Fe) &   Be &   -- & 190      & -- & 4.26 & 12 &	      \\
NGC\,346-074 & B3 III       & 16500& 3.20 &  52      & He & 3.80 &  7 &	      \\
NGC\,346-075 & B1 V         & 27700& 4.30 &  10      & H06& 4.31 & 13 & Binary\\
NGC\,346-076 & B2 (Be-Fe)   &   Be &   -- & 237      & -- & 4.21 & 11 &	      \\
NGC\,346-077 & O9 V         & 36500& 3.99 & 177      & M06& 4.65 & 20 &	      \\
NGC\,346-078 & B2 III       &  SB2?&   -- & 154      & A  & 4.05 &  9 &Binary?\\
NGC\,346-079 & B0.5 Vn      & 29500& 4.20 & 293      & He & 4.37 & 14 &	      \\
NGC\,346-080 & B1 V         & 27300& 4.25 & 216      & A  & 4.29 & 12 &	      \\
NGC\,346-081 & B2 IIIn      & 21200& 3.50 & 255      & A  & 4.03 &  9 &	      \\
NGC\,346-082 & B2 III       & 21200& 3.70 & 168      & A  & 4.03 &  9 & Binary\\
NGC\,346-083 & B1 V         & 27300& 4.05 & 207      & A  & 4.28 & 12 &	      \\
NGC\,346-084 & B1 V         & 27300& 4.25:& 105      & A  & 4.27 & 12 &	      \\
NGC\,346-085 & B2 III       &  SB2 &   -- &  26      & -- & 4.03 &  9 & Binary\\
NGC\,346-085B&              &  SB2 &   -- &  11      & -- &   -- & -- & Binary\\
NGC\,346-088 & B1 V         & 27300& 4.10:&  84      & A  & 4.26 & 12 &	      \\
NGC\,346-089 & B1-2 (Be-Fe) &   Be &   -- &  79      & -- & 4.21 & 11 &	      \\
NGC\,346-090 & O9.5 V       & 34900& 4.25 & 188      & M06& 4.56 & 18 &	      \\
NGC\,346-091 & B1e          &   Be &   -- &   49     & -- & 4.25 & 12 & Binary\\
NGC\,346-092 & B1 Vn        & 27300& 3.95 &  234     & A  & 4.24 & 12 &       \\
NGC\,346-093 & B0 V         & 34400& 4.42 &  187     & M06& 4.53 & 18 &       \\
NGC\,346-094 & B0.7 V       & 28500& 4.00 &   40     & H06& 4.28 & 13 &       \\
NGC\,346-095 & B1-2 (Be-Fe) &   Be &   -- &  227     & -- & 4.18 & 11 &       \\
NGC\,346-096 & B1-2 (Be-Fe) &   Be &   -- &  343     & -- & 4.18 & 11 &       \\
NGC\,346-097 & O9 V         & 37500& 4.49 &   22     & M06& 4.75 & 22 &       \\
NGC\,346-098 & B1.5 V       & 26100& 4.05 &   56     & A  & 4.16 & 11 &       \\
NGC\,346-099 & B3 III       & 18000& 3.40 &   94     & He & 3.79 &  7 &       \\
NGC\,346-100 & B1.5 V       & 26100& 4.30 &  183     & A  & 4.16 & 11 &       \\
NGC\,346-101 & B1 V         & 27300& 4.25 &   29     & A  & 4.18 & 12 &       \\
NGC\,346-102 & B3 III       & 17700& 3.70 &   30     & He & 3.75 &  7 &       \\
NGC\,346-103 & B0.5 V       & 29500& 4.00 &   10     & H06& 4.26 & 13 &       \\
NGC\,346-104 & B0 V         & 33500& 4.45 &   17     & He & 4.41 & 16 & Binary\\
NGC\,346-106 & B1 V         & 27500& 4.20 &  142     & He & 4.17 & 12 & Binary\\
NGC\,346-107 & O9.5 V       & 35900& 4.23 &   55     & M06& 4.40 & 18 &       \\
NGC\,346-108 & B1.5 V       & 26100& 3.95 &  167     & A  & 4.11 & 11 &       \\
NGC\,346-109 & B1.5 V       & 26100& 4.25 &  123     & A  & 4.11 & 11 &       \\
NGC\,346-110 & B1-2 (Be-Fe) &   Be &   -- &  243     & -- & 4.11 & 11 & Binary\\
NGC\,346-112 & O9.5 V       & 34400& 4.15 &  143     & M06& 4.36 & 17 &       \\
NGC\,346-114 & B1 Vn        & 27300& 3.90 &  287     & A  & 4.13 & 12 &       \\
NGC\,346-116 & B1 V         & 28250& 4.10 &   15     & H06& 4.15 & 12 &       \\
\\
\hline
\end{tabular}
\end{table*}

\begin{table*}
\caption{Atmospheric parameters and $v\sin i$ values of the NGC\,330 sample. 
Symbols and labels are the same as given in
Table~\ref{t_346_vsini}.} 
\label{t_330_vsini}
\centering
\begin{tabular}{llcccccccl}\hline \hline
Star&Spectral Type&T$_{\rm eff}$&$\log g$&$v\sin i$&Method&L/L$_{\rm \sun}$&Mass
                                                                    &Comments \\
             &              & (K)&(dex) &(km\,s$^{-1})$&&  &(M$_{\rm \sun}$)&        \\
\hline \\
NGC\,330-002 & B3 Ib        &14590 &2.15  &$<$24     & T06& 4.77 & 15 &	      \\
NGC\,330-003 & B2 Ib        &17210 &2.25  &$<$36     & T06& 4.89 & 16 & Binary\\
NGC\,330-004 & B2.5 Ib      &17000 &2.30  &$<$41     & T06& 4.81 & 15 &	      \\
NGC\,330-005 & B5 Ib        &13700 &2.25  &  20      & T06& 4.58 & 13 &	      \\
NGC\,330-009 & B5 Ib	    &13940 &2.45  &  28      & T06& 4.45 & 12 &	      \\
NGC\,330-010 & B5 Ib	    &14820 &2.60  &$<$15     & T06& 4.44 & 12 &	      \\
NGC\,330-013&O8.5 III-II((f))&34500&3.40  &  78      & M06& 5.40 & 32 &	      \\
NGC\,330-014 & B1.5 Ib      &20130 &2.75  &  45      & T06& 4.69 & 14 &	      \\
NGC\,330-016 & B5: II	    &14220 &2.60  &  40      & T06& 4.24 & 10 &	      \\
NGC\,330-017 & B2 II	    &22000 &3.35  &  21      & T06& 4.66 & 14 &	      \\
NGC\,330-018 & B3 II	    &18000 &2.95  &  48      & T06& 4.45 & 12 &	      \\
NGC\,330-020 & B3 II	    &16720 &2.85  &$<$42     & T06& 4.35 & 11 &	      \\
NGC\,330-021 & B0.2 III     &30500 &3.70  & 204      & He & 4.95 & 21 &	      \\
NGC\,330-022 & B3 II	    &18860 &3.00  &  24      & T06& 4.42 & 12 &	      \\
NGC\,330-024 & B5 Ib	    &14000 &2.70  &  46      & He & 4.05 &  9 &	      \\
NGC\,330-025 & B1.5e	    &	Be &  --  & 178      & -- & 4.64 & 15 &	      \\
NGC\,330-026 & B2.5 II      &22500 &3.40  &  73      & T06& 4.50 & 12 &	      \\
NGC\,330-027 & B1 V	    &22040 &3.20  &  80      & T06& 4.46 & 12 &	      \\
NGC\,330-028 & B1 V	    &27300 &3.75  &  76      & A  & 4.63 & 15 &	      \\
NGC\,330-029 &B0.2 V(Be-Fe) &   Be &  --  & 209      & -- & 4.75 & 18 & Binary\\
NGC\,330-031 &B0.5 V(Be-Fe) &   Be &  --  & 178      & -- & 4.69 & 17 &	      \\
NGC\,330-032 & B0.5 V	    &29700 &4.15  &  17      & T06& 4.67 & 17 &	      \\
NGC\,330-033 & B1.5 V	    &26100 &3.90  & 105      & A  & 4.53 & 14 &	      \\
NGC\,330-034 & B1-2e        &   Be &  --  & 231      & -- & 4.53 & 14 &	      \\
NGC\,330-035 & B3 II	    &19000 &3.20  &  40      & He & 4.22 & 10 &	      \\
NGC\,330-036 & B2 II	    &21200 &3.25  &  39      & A  & 4.33 & 11 &	      \\
NGC\,330-038 & B1 V	    &27300 &3.75  & 150      & A  & 4.53 & 14 &	      \\
NGC\,330-039 & B0 V	    &33000 &4.25  & 120      & He & 4.72 & 18 &	      \\
NGC\,330-040 & B2 III	    &21200 &3.25  & 106      & A  & 4.25 & 11 &	      \\
NGC\,330-041 & B0 V	    &32000 &4.15: & 127      & He & 4.64 & 17 &	      \\
NGC\,330-042 & B2 II	    &25450 &3.75  &  27      & T06& 4.38 & 12 &	      \\
NGC\,330-043 & B0 V	    &33000 &4.10: & 250      & He & 4.64 & 18 &	      \\
NGC\,330-044 & B1-2 (Be-Fe) &   Be &  --  & 184      & -- & 4.37 & 12 &	      \\
NGC\,330-045 & B3 III	    &18450 &3.25  & 133      & A  & 4.01 &  9 & Binary\\
NGC\,330-046 & O9.5 V	    &34000 &4.25: & 165      & He & 4.64 & 18 &	      \\
NGC\,330-047 & B1 V	    &26700 &4.00  &  28      & T06& 4.35 & 12 &	      \\
NGC\,330-048 & B0.5 V	    &29000 &4.00  &  73      & He & 4.42 & 14 &	      \\
NGC\,330-049 & O9 V	    &35000 &4.15  &  40      & He & 4.64 & 19 &	      \\
NGC\,330-050 & B3e	    &	Be &  --  & 214      & -- & 4.05 &  9 &	      \\
NGC\,330-051 & B1.5 V	    &26100 &4.05  & 273      & A  & 4.30 & 12 &	      \\
NGC\,330-052 & O8.5 Vn      &35700 &3.91  & 291      & M06& 4.60 & 19 &	      \\
NGC\,330-053 & B0.5 V	    &29650 &4.25  &  82      & A  & 4.42 & 14 & Binary\\
NGC\,330-054 & B2 (Be-Fe)   &   Be &  --  & 147      & -- & 4.25 &  8 &	      \\
NGC\,330-055 & B0.5 V	    &31000 &4.10  & 219      & He & 4.46 & 15 &Binary?\\
NGC\,330-056 & B2 III	    &21200 &3.50  & 108      & A  & 4.06 &  9 &	      \\
NGC\,330-057 & B0.5 V	    &29000 &4.15  & 104      & He & 4.34 & 13 &	      \\
NGC\,330-058 & B3:	    &  SP? &  --  & 263      & -- & 3.97 & 10 &	      \\
NGC\,330-059 & B3 III	    &18450 &3.25  & 123      & A  & 3.89 &  8 &	      \\
NGC\,330-060 & B2.5 (Be-Fe) &	Be &  --  &  88      & -- & 4.08 &  9 &	      \\
NGC\,330-062 & B3e	    &	Be &  --  & 241      & -- & 3.95 &  8 &	      \\
NGC\,330-063 & B1-3	    &  SP? &  --  & 199      & -- & 4.17 & 11 &	      \\
NGC\,330-064 & B3:e         &   Be &  --  & 269      & -- & 3.93 &  8 &	      \\
NGC\,330-065 & B1-3 (Be-Fe) &	Be &  --  & 284      & -- & 4.13 & 11 &	      \\
NGC\,330-066 & B3 III	    &18500 &3.40  & 126      & He & 3.83 &  8 &	      \\
NGC\,330-067 & B2.5 III     &19850 &3.40  &  65      & A  & 3.90 &  8 &	      \\
\\
\hline
\end{tabular}
\end{table*}

\begin{table*}
\addtocounter{table}{-1}
\caption{-continued.}
\centering
\begin{tabular}{llcccccccl}\hline \hline
Star&Spectral Type&T$_{\rm eff}$&$\log g$&$v\sin i$&Method&L/L$_{\rm \sun}$&Mass
                                                                    &Comments \\
             &              & (K)&(dex) &(km\,s$^{-1})$&&  &(M$_{\rm \sun}$)&        \\
\hline \\
NGC\,330-068 & B1.5 (Be-Fe) &	Be &  --  &  23      & -- & 4.16 & 11 &	      \\
NGC\,330-069 & B3 III	    &19000 &3.40  & 193      & He & 3.85 &  8 &	      \\
NGC\,330-070 & B0.5e	    &	Be &  --  & 348      & -- & 4.29 & 13 &	      \\
NGC\,330-071 & B3 III	    &19000 &3.50  &  92      & He & 3.84 &  8 &	      \\
NGC\,330-072 & B0.5 V	    &29650 &4.15  &  99      & A  & 4.29 & 13 &	      \\
NGC\,330-073 & B8 Ib	    &12000 &2.20  &  49      & He & 3.37 &  5 &	      \\
NGC\,330-074 & B0 V	    &32020 &4.20  &  29      & T06& 4.35 & 15 &	      \\
NGC\,330-075 & B8 II	    &13500 &3.20  & 166:     & He & 3.46 &  6 &	      \\
NGC\,330-076 & B3 (Be-Fe)   &	Be &  --  &  62      & -- & 3.86 &  8 &	      \\
NGC\,330-079 & B3 III	    &19500 &3.40  & 146      & He & 3.82 &  8 & Binary\\
NGC\,330-080 & B1-3	    &  SP? &  --  & 283      & -- & 4.06 & 10 &	      \\
NGC\,330-081 & B1-3	    &19000 &3.50  & 190      & He & 3.79 &  7 &	      \\
NGC\,330-082 & B1-3	    &  SP? &  --  & 192      & -- & 4.05 & 10 &	      \\
NGC\,330-083 & B3 III	    &18000 &3.25  & 140      & He & 3.72 &  7 &	      \\
NGC\,330-084 & B3 V-III     & SB2? &  --  &  26      & -- & 3.83 &  8 & Binary\\
NGC\,330-085 & B3:e         &	Be &  --  & 191      & -- & 3.82 &  8 &	      \\
NGC\,330-086 & B2.5 III     &19850 &3.50  &  89      & A  & 3.80 &  8 &	      \\
NGC\,330-087 & Be-Fe	    &	Be &  --  & 214      & -- & 3.81 &  8 &	      \\
NGC\,330-089 & B1-5	    &18500 &3.50  & 193      & He & 3.70 &  7 &	      \\
NGC\,330-090 & B3 III	    &18450 &3.30  & 167      & A  & 3.69 &  7 &	      \\
NGC\,330-091 & B0e	    &	Be &  --  & 272:     & -- & 4.25 & 14 &	      \\
NGC\,330-094 & B1-5	    &  SP? &  --  & 330      & -- & 3.77 &  8 &	      \\
NGC\,330-095 & B3 III	    &18450 &3.45  &  20      & A  & 3.67 &  7 &	      \\
NGC\,330-096 & B1-3 (Be-Fe) &	Be &  --  & 175:     & -- & 3.97 & 10 &	      \\
NGC\,330-097 & B1 V	    &27300 &3.90  & 154      & A  & 4.05 & 11 &	      \\
NGC\,330-098 & B0.2: V      &  SP? &  --  & 148      & -- & 4.18 & 13 &	      \\
NGC\,330-099 & B2-3	    &  SP? &  --  & 187      & -- & 3.86 &  9 &	      \\
NGC\,330-100 & Be (B0-3)    &  Be  &  --  & 373      & -- & 3.99 & 10 &	      \\
NGC\,330-101 & B2.5 III     &19850 &3.50  &  48      & A  & 3.73 &  7 &	      \\
NGC\,330-102 & B2-3 III     &19850 &3.60  & 129      & A  & 3.72 &  7 &	      \\
NGC\,330-103 & B1-3	    &  SP? &  --  & 131      & -- & 3.94 &  9 &	      \\
NGC\,330-104 & B0: V	    &  SP? &  --  & 309:     & -- & 4.19 & 14 &	      \\
NGC\,330-105 & B1-3	    &  SP? &  --  &  13      & -- & 3.91 &  9 &	      \\
NGC\,330-106 & B1-2	    &  SP? &  --  &  71      & -- & 3.95 & 10 &	      \\
NGC\,330-107 & B3: V-III    &16500 &3.25  &  34      & He & 3.49 &  6 & Binary\\
NGC\,330-108 & B5 III	    &14500 &3.30  & 102      & He & 3.35 &  5 &	      \\
NGC\,330-109 & B3 III	    &18450 &3.20  & 213      & A  & 3.60 &  7 &	      \\
NGC\,330-110 & B2 III	    &21200 &3.40  & 110      & A  & 3.74 &  8 &	      \\
NGC\,330-111 & B1-3	    &  SP? &  --  & 173      & -- & 3.79 &  8 &	      \\
NGC\,330-112 & B1-3 (Be-Fe) &  Be  & -- - & 262      & -- & 3.79 &  8 &	      \\
NGC\,330-113 & B1-3	    &  SP? &  --  & 327      & -- & 3.78 &  8 &	      \\
NGC\,330-114 & B2 III	    &23800 &3.90  &  17      & T06& 3.83 &  9 &	      \\
NGC\,330-116 & B3 III	    &19500 &3.55  &   4      & He & 3.62 &  7 &	      \\
NGC\,330-118 & B1-2	    &  SP? &  --  &  32      & -- & 3.89 & 10 &	      \\
NGC\,330-119 & B1-3	    &  SP? &  --  & 265      & -- & 3.84 &  9 &	      \\
NGC\,330-120 & B3: V-III    &18500 &3.40  & 137:     & He & 3.47 &  6 &	      \\
NGC\,330-123 & O9.5 V	    &35000 &4.25: &  26      & He & 4.58 & 18 &	      \\
NGC\,330-124 & B0.2 V	    &30980 &4.25  &  95      & T06& 4.42 & 15 &	      \\
NGC\,330-125 & B2 III	    &21200 &3.50  &  86      & A  & 4.01 &  9 &	      \\
\\
\hline
\end{tabular}
\end{table*}

\begin{table*}
\caption{Atmospheric parameters and $v\sin i$ values of the N\,11 sample. 
Symbols and labels are the same as given in
Table~\ref{t_346_vsini}.} 
\label{t_11_vsini}
\centering
\begin{tabular}{llcccccccl}\hline \hline
Star&Spectral Type&T$_{\rm eff}$&$\log g$&$v\sin i$&Method&L/L$_{\rm \sun}$&Mass
                                                                    &Comments \\
             &              & (K)&(dex) &(km\,s$^{-1})$&&  &(M$_{\rm \sun}$)&        \\
\hline \\
N\,11-001    & B2 Ia        &18750 &2.50  &$<$52     & H06& 5.66 & 38 &       \\
N\,11-002    & B3 Ia        &15800 &2.10  &$<$53     & H06& 5.26 & 24 &       \\
N\,11-003    & B1 Ia        &23200 &2.75  &$<$54     & H06& 5.42 & 30 &       \\
N\,11-004    & O9.7 Ib      &31600 &3.36  &  40      & M06& 5.80 & 48 &       \\
N\,11-008    & B0.5 Ia      &25450 &3.00  &$<$43     & H06& 5.39 & 30 &       \\
N\,11-009    & B3 Iab       &15000 &2.15  &  30      & H06& 4.85 & 17 & Binary\\
N\,11-010 & O9.5 III +B1-2  &  SB2 &  --  & 152      &  --& 5.59 & 40 & Binary\\
N\,11-010B   &              &  SB2 &  --  &  12      &  --&   -- & -- & Binary\\
N\,11-011    & OC9.5 II     &29500 &3.15  &$<$60     &  He& 5.49 & 35 & Binary\\
N\,11-012    & B1 Ia        &20500 &2.55  &  40      & H06& 5.13 & 22 &       \\
N\,11-014    & B2 Iab       &19100 &2.55  &  30      & H06& 5.03 & 19 & Binary\\
N\,11-015    & B0.7 Ib      &23600 &2.95  &  58      & H06& 5.23 & 24 &       \\
N\,11-016    & B1 Ib        &21700 &2.75  &  54      & H06& 5.13 & 22 &       \\
N\,11-017    & B2.5 Iab     &16500 &2.30  &$<$25     & H06& 4.82 & 17 & Binary\\
N\,11-023    & B0.7 Ib      &24000 &2.90  &  66      & H06& 5.09 & 21 &       \\
N\,11-024    & B1 Ib        &21600 &2.80  &$<$30     & H06& 4.96 & 18 &       \\
N\,11-026    & O2.5 III(f*) &53300 &4.00  & 109      & M06& 5.92 & 82 &       \\
N\,11-029    & OC9.7 Ib     &28750 &3.30  &  40      & H06& 5.21 & 25 &       \\
N\,11-031    & ON2 III(f*)  &45000 &3.85  & 116      & M06& 5.84 & 61 &       \\
N\,11-032    & O7 II(f)     &35200 &3.45  & $<$65    & M06& 5.43 & 34 &       \\
N\,11-033    & B0 IIIn      &27200 &3.21  & 256      & M06& 5.07 & 21 &       \\
N\,11-034    & B0.5 III     &25500 &3.25  & 203      &  He& 5.03 & 20 & Binary\\
N\,11-035    & O9 II(f)     &31000 &3.25  &$<$56     &  He& 5.23 & 27 & Binary\\
N\,11-036    & B0.5 Ib      &23750 &3.10  &  30      & H06& 4.95 & 18 &       \\
N\,11-037    & B0 III       &28100 &3.25  & 100      & H06& 5.08 & 23 & Binary\\
N\,11-038    & O5 III(f+)   &41000 &3.72  & 145      & M06& 5.69 & 48 &       \\
N\,11-039    & B2 III       &21700 &3.00  & 157      &  A & 4.81 & 16 &Binary?\\
N\,11-040    & B0: IIIn     &29500 &3.30  & 273      &  He& 5.11 & 24 &       \\
N\,11-042    & B0 III       &29000 &3.60  &  30      & H06& 5.05 & 22 & Binary\\
N\,11-045    & O9-9.5 III   &32300 &3.32  & 105      & M06& 5.15 & 25 &       \\
N\,11-046    & O9.5 V       &33500 &4.25  & 205      &  He& 5.21 & 28 & Binary\\
N\,11-047    & B0 III       &29200 &3.65  &  55      & H06& 5.03 & 22 & Binary\\
N\,11-048    & O6.5 V((f))  &40700 &4.19  & 130      & M06& 5.38 & 37 &       \\
N\,11-051    & O5 Vn((f))   &42400 &3.75  & 333      & M06& 5.31 & 36 &       \\
N\,11-052    & O9.5 V       &  SB2 &	  &  16      &  --& 5.23 & 29 & Binary\\
N\,11-054    & B1 Ib        &23500 &3.05  &  45      & H06& 4.79 & 16 &       \\
N\,11-056    & B1e          &Be/SB2&  --  & 205      &  --& 4.89 & 19 & Binary\\
N\,11-056B   &              &Be/SB2&  --  &  25      &  --&  --  & -- & Binary\\
N\,11-058    & O5.5 V((f))  &41300 &3.89  &  85      & M06& 5.27 & 34 &       \\
N\,11-059    & O9 V         &34000 &3.85  & 103      &  He& 5.13 & 26 & Binary\\
N\,11-060    & O3 V ((f*))  &45700 &3.92  & 106      & M06& 5.57 & 49 &       \\
N\,11-061    & O9 V         &33600 &3.51  &  87      & M06& 5.20 & 27 &       \\
N\,11-062    & B0.2 V       &30400 &4.05  &  25      & H06& 4.95 & 21 & Binary\\
N\,11-063    & O9: V        &35000 &4.25  & 196      &  He& 5.11 & 26 &       \\
N\,11-064    & B0.2: Vn     &  SB2?&  --  & 206      &  --& 4.92 & 20 &Binary?\\
N\,11-065    & O6.5 V((f))  &41700 &3.89  &  83      & M06& 5.17 & 33 &       \\
N\,11-066    & O7 V((f))    &39300 &3.87  &  71      & M06& 5.10 & 30 &       \\
N\,11-068    & O7 V((f))    &39900 &4.13  &  54      & M06& 5.06 & 29 &       \\
N\,11-069    & B1 III       &24300 &3.30  &  80      & H06& 4.63 & 15 &       \\
N\,11-070    & B3 III       &19500 &3.30  &  62      &  He& 4.40 & 12 &       \\
N\,11-072    & B0.2 V       &28800 &3.75  &  15      & H06& 4.77 & 18 &       \\
N\,11-073    & B0.5 (Be-Fe) &   Be &  --  & 222      &  --& 4.78 & 18 &       \\
N\,11-074    & B0.5 (Be-Fe) &   Be &  --  & 169      &  --& 4.78 & 18 &       \\
N\,11-075    & B2 III       &21800 &3.35  &  25      & H06& 4.48 & 12 & Binary\\
N\,11-076    & B0.2 Ia      &26500 &2.90  &  72      &  He& 4.67 & 16 &       \\
\\
\hline
\end{tabular}
\end{table*}

\begin{table*}
\addtocounter{table}{-1}
\caption{-continued.}
\centering
\begin{tabular}{llcccccccl}\hline \hline
Star&Spectral Type&T$_{\rm eff}$&$\log g$&$v\sin i$&Method&L/L$_{\rm \sun}$&Mass
                                                                    &Comments \\
             &              & (K)&(dex) &(km\,s$^{-1})$&&  &(M$_{\rm \sun}$)&        \\
\hline \\
N\,11-077    & B2 III       &21500 &3.30  & 117      &  He& 4.47 & 12 & Binary\\
N\,11-078    & B2 (Be-Fe)   &   Be &  --  &  93      &  --& 4.59 & 14 & Binary\\
N\,11-079    & B0.2 V       &32500 &4.25  &  38      &  He& 4.88 & 21 &       \\
N\,11-081    & B0:n (Be-Fe) &   Be &  --  & 363      &  He& 4.79 & 19 &       \\
N\,11-082    & B1-2 +early B&  SB2 &  --  &  95      &  --& 4.58 & 14 & Binary\\
N\,11-082B   &              &  SB2 &  --  & 110      &  --&   -- & -- & Binary\\
N\,11-083    & B0.5 V       &29300 &4.15  &  20      & H06& 4.71 & 17 & Binary\\
N\,11-084    & B0.5 V       &29500 &4.25  & 108      &  He& 4.75 & 18 & Binary\\
N\,11-085    & B0.5 V       &  SB2 &  --  & 105      &  --& 4.73 & 17 & Binary\\
N\,11-085B   &              &  SB2 &  --  &  28      &  --&   -- & -- & Binary\\
N\,11-086    & B1 V         &26800 &4.25  &  75      &  A & 4.65 & 16 &       \\
N\,11-087    & O9.5 Vn      &32700 &4.04  & 276      & M06& 4.91 & 21 &       \\
N\,11-088    & B1 III       &24150 &3.45  & 240      &  A & 4.54 & 14 &       \\
N\,11-089    & B2 III       &21700 &3.20  & 117      &  A & 4.42 & 12 & Binary\\
N\,11-090    & B2e          &   Be &  --  & 104      &  --& 4.53 & 14 &       \\
N\,11-093    & B2.5 III     &19500 &3.30  &  73      &  He& 4.29 & 11 &       \\
N\,11-094    & B1 III       &  SB2 &  --  & 147      &  --& 4.50 & 13 & Binary\\
N\,11-095    & B1 Vn        &26800 &3.85  & 267      &  A & 4.59 & 15 &       \\
N\,11-096    & B1 III       &   SB2&  --  &  32      &  --& 4.49 & 13 & Binary\\
N\,11-097    & B3 II        &17500 &2.95  &  52      &  He& 4.17 & 10 &       \\
N\,11-098    & B2 III       &21700 &3.50  &  45      &  A & 4.37 & 12 & Binary\\
N\,11-100    & B0.5 V       &29700 &4.15  &  30      & H06& 4.68 & 17 &       \\
N\,11-101    & B0.2 V       &29800 &3.95  &  70      & H06& 4.68 & 17 &       \\
N\,11-102    & B0.2 V       &31000 &4.20  & 218      &  He& 4.73 & 18 &       \\
N\,11-103    &B1-2 + early B&  SB2 &  --  & 125      &  --& 4.53 & 14 & Binary\\
N\,11-103B   &              &  SB2 &  --  &  80      &  --&   -- & -- & Binary\\
N\,11-104    & B1.5 V       &25700 &3.75  & 153      &  A & 4.52 & 14 &       \\
N\,11-105    & B1 V         & SB2? &  --  & 116      &  --& 4.56 & 15 &Binary?\\
N\,11-106    & B0 V	    &31200 &4.00  &  25      & H06& 4.72 & 18 &       \\
N\,11-107    & B1-2 + early B&  SB2&  --  &  80      &  --& 4.51 & 13 & Binary\\
N\,11-107B   &              &  SB2 &  --  & 170      &  --&   -- & -- & Binary\\
N\,11-108    & O9.5 V       &32150 &4.10  &  25      & H06& 4.73 & 19 &       \\
N\,11-109    & B0.5 Ib      &25750 &3.20  &  40      & H06& 4.48 & 12 &       \\
N\,11-110    & B1 III       &23100 &3.25  &  25      & H06& 4.37 & 12 &       \\
N\,11-111    & B1.5 III     &22950 &3.50  & 101      &  A & 4.36 & 12 &       \\
N\,11-113    & B0.5 III	    &  SB2 &  --  &  38      &  --& 4.49 & 14 & Binary\\
N\,11-113B   &              &  SB2 & --   &  20      &  --& --   & -- & Binary\\
N\,11-114    & B0 Vn        &32500 &4.05  & 299      &  He& 4.70 & 19 &       \\
N\,11-115    & B1 III       &24150 &3.65  &  53      &  A & 4.39 & 12 &       \\
N\,11-116    & B2 III       &21700 &3.30  & 160      &  A & 4.28 & 11 &       \\
N\,11-117    & B1 Vn        &26800 &4.20  & 247      &  A & 4.46 & 13 &       \\
N\,11-118    & B1.5 V       &25700 &3.80  & 150      &  A & 4.42 & 13 & Binary\\
N\,11-119    & B1.5 V       &  SB2 &  --  & 259      &  --& 4.42 & 13 & Binary\\
N\,11-120    & B0.2 Vn      &31500 &4.10  & 207      &  He& 4.63 & 17 &       \\
N\,11-121    & B1 Vn        &26800 &4.20  & 265      &  A & 4.45 & 14 &       \\
N\,11-122    & O9.5 V       &33000 &4.10  & 173      &  He& 4.62 & 18 &       \\
N\,11-123    & O9.5 V       &34800 &4.22  & 110      & M06& 4.58 & 19 &       \\
N\,11-124    & B0.5 V       &28500 &4.20  &  45      & H06& 4.47 & 14 & Binary\\
\\
\hline
\end{tabular}
\end{table*}

\begin{table*}
\caption{Atmospheric parameters and $v\sin i$ values of the NGC\,2004 sample. 
Symbols and labels are the same as given in
Table~\ref{t_346_vsini}.} 
\label{t_2004_vsini}
\centering
\begin{tabular}{llcccccccl}\hline \hline
Star&Spectral Type&T$_{\rm eff}$&$\log g$&$v\sin i$&Method&L/L$_{\rm \sun}$&Mass
                                                                    &Comments \\
             &              & (K)&(dex) &(km\,s$^{-1})$&&  &(M$_{\rm \sun}$)&        \\
\hline \\
NGC\,2004-003 & B5 Ia	    &14450 &2.10 &  33      & T06& 5.04 & 20  & Binary\\
NGC\,2004-005 & B8 Ia	    &12390 &1.90 &  30      & T06& 4.95 & 18  & Binary\\
NGC\,2004-007 & B8 IIa	    &12250 &2.00 &$<$30     & T06& 4.90 & 17  &       \\
NGC\,2004-010 & B2.5 Iab    &17160 &2.40 &$<$30     & T06& 5.02 & 19  &       \\
NGC\,2004-011 & B1.5 Ia     &21250 &2.75 &$<$40     & T06& 5.22 & 24  &       \\
NGC\,2004-012 & B1.5 Iab    &21270 &2.87 &$<$40     & T06& 4.92 & 18  &       \\
NGC\,2004-013 & B2 II	    &21700 &3.20 & 145      & A  & 4.94 & 18  &       \\
NGC\,2004-014 & B3 Ib	    &17800 &2.85 &$<$30     & T06& 4.72 & 15  &       \\
NGC\,2004-015 & B1.5 II     &22950 &3.15 & 165      & A  & 4.96 & 18  & Binary\\
NGC\,2004-020 & B1.5 II     &22950 &3.20 & 145      & A  & 4.91 & 18  & Binary\\
NGC\,2004-021 & B1.5 Ib     &21450 &3.00 &$<$40     & T06& 4.82 & 16  &       \\
NGC\,2004-022 & B1.5 Ib     &21780 &3.15 &$<$30     & T06& 4.79 & 16  &       \\
NGC\,2004-023 & B2 (Be-Fe)  &  Be  &  -- & 102:     & -- & 4.85 & 17  &Binary?\\
NGC\,2004-024 & B1.5 IIIn   &22950 &3.05 & 240      & A  & 4.72 & 15  &       \\
NGC\,2004-025 & B2 (Be-Fe)  &	Be &  -- &  83      & -- & 4.75 & 15  & Binary\\
NGC\,2004-026 & B2 II	    &22900 &3.65 &$<$20     & T06& 4.68 & 15  & Binary\\
NGC\,2004-027 & B0e	    &	Be &  -- & 182:     & -- & 5.00 & 23  &       \\
NGC\,2004-029 & B1.5e	    &23100 &3.50 &  30      & T06& 4.65 & 14  & Binary\\
NGC\,2004-030 & B0.2 Ib     &29000 &3.75 & 123      & He & 4.87 & 19  & Binary\\
NGC\,2004-031 & B2 II       &21700 &3.35 &  65      & A  & 4.58 & 13  & Binary\\
NGC\,2004-032 & B2 II       &21700 &3.40 & 110      & A  & 4.58 & 13  &       \\
NGC\,2004-035 & B1: (Be-Fe) &	Be &  -- & 244:     & -- & 4.74 & 17  &       \\
NGC\,2004-036 & B1.5 III    &22870 &3.35 &  42      & T06& 4.58 & 13  &       \\
NGC\,2004-039 & B1.5e	    &	Be &  -- & 212:     & -- & 4.67 & 15  &       \\
NGC\,2004-041 & B2.5 III    &20450 &3.30 & 101      & A  & 4.43 & 12  & Binary\\
NGC\,2004-042 & B2.5 III    &20980 &3.45 &  42      & T06& 4.45 & 12  &       \\
NGC\,2004-043 & B1.5 III    &22950 &3.50 &  24      & A  & 4.52 & 13  &       \\
NGC\,2004-044 & B1.5:	    &  SB2 & --  &  18      & -- & 4.62 & 15  & Binary\\
NGC\,2004-044B&             &  SB2 & --  &  55      & -- &  --  & --  & Binary\\
NGC\,2004-045 & B2 III      &21700 &3.50 & 128      & A  & 4.46 & 12  & Binary\\
NGC\,2004-046 & B1.5 III    &26090 &3.85 &  32      & T06& 4.62 & 15  &       \\
NGC\,2004-047 & B2 III      &21700 &3.35:& 133      & A  & 4.42 & 12  & Binary\\
NGC\,2004-048 & B2.5e	    &   Be &  -- & 173      & -- & 4.49 & 13  &       \\
NGC\,2004-050 & B2.5 III    &20450 &3.30 & 109      & A  & 4.35 & 11  & Binary\\
NGC\,2004-051 & B2 III      &21700 &3.40 &  70      & A  & 4.40 & 12  &       \\
NGC\,2004-052 & B2 III      &21700 &3.60 & 138      & A  & 4.40 & 12  &       \\
NGC\,2004-053 & B0.2 Ve     &31500 &4.15 &   7      & T06& 4.77 & 18  &       \\
NGC\,2004-054 & B2 III      &21700 &3.40 & 114      & A  & 4.39 & 12  & Binary\\
NGC\,2004-055 & B2.5 III    &20450 &3.30 & 118      & A  & 4.33 & 11  &       \\
NGC\,2004-056 & B1.5e	    &	Be &  -- & 229      & -- & 4.52 & 14  &       \\
NGC\,2004-058 &O9.5 V (Nstr)&33500 &4.10 & 195      & He & 4.80 & 20  & Binary\\
NGC\,2004-059 & B2 III      &21700 &3.45 &  91      & A  & 4.35 & 11  & Binary\\
NGC\,2004-060 & B2 III      &21700 &3.45 & 134      & A  & 4.35 & 11  &       \\
NGC\,2004-061 & B2 III      &20990 &3.35 &  40      & T06& 4.31 & 11  &       \\
NGC\,2004-062 & B0.2 V      &30250 &4.35 & 106      & A  & 4.67 & 17  &       \\
NGC\,2004-063 & B2 III      &21700 &3.50 & 107      & A  & 4.32 & 11  &       \\
NGC\,2004-064 & B0.7-B1 III &25900 &3.70 &  28      & T06& 4.48 & 13  &       \\
NGC\,2004-065 & B2.5 III    &20450 &3.40 & 223      & A  & 4.25 & 11  &Binary?\\
NGC\,2004-066 & B1.5 Vn     &25700 &3.70 & 238      & A  & 4.48 & 13  &       \\
NGC\,2004-067 & B1.5e	    &	Be &  -- & 237      & -- & 4.48 & 13  &       \\
NGC\,2004-068 & B2.5 III    &20450 &3.65 &  62      & A  & 4.25 & 11  &       \\
NGC\,2004-069 & B0.7 V      &28000 &4.00 & 178      & He & 4.54 & 15  &       \\
NGC\,2004-070 & B0.7-B1 III &27400 &3.90 &  46      & T06& 4.51 & 14  &       \\
NGC\,2004-071 & B1.5 III    &22950 &3.50 &  98      & A  & 4.33 & 12  &       \\
NGC\,2004-073 & B2 III      &21700 &3.55 &  37      & A  & 4.26 & 11  &       \\
\\
\hline
\end{tabular}
\end{table*}

\begin{table*}
\addtocounter{table}{-1}
\caption{-continued.}
\centering
\begin{tabular}{llcccccccl}\hline \hline
Star&Spectral Type&T$_{\rm eff}$&$\log g$&$v\sin i$&Method&L/L$_{\rm \sun}$
                                                              &Mass &Comments \\
             &              & (K)&(dex) &(km\,s$^{-1})$&&  &(M$_{\rm \sun}$)
	                                                             &        \\
\hline \\
NGC\,2004-074 & B0.7-B1 V   &27375 &4.25 & 130      & A  & 4.49 & 14  & Binary\\
NGC\,2004-075 & B2 III      &21700 &3.55 & 116      & A  & 4.26 & 11  &       \\
NGC\,2004-076 & B2.5 III    &20450 &3.65 &  37      & A  & 4.20 & 10  &       \\
NGC\,2004-077 & B0.5 V      &29500 &4.00:& 215      & He & 4.56 & 16  &       \\
NGC\,2004-078 & B2 III      &21700 &3.55 & 115      & A  & 4.26 & 11  & Binary\\
NGC\,2004-079 & B2 III      &21700 &3.60 & 165      & A  & 4.25 & 11  & Binary\\
NGC\,2004-080 & B2.5 III    &20450 &3.40 &  85      & A  & 4.19 & 10  &       \\
NGC\,2004-081 & B1 V	    &26800 &4.00 & 105:     & A  & 4.45 & 13  &       \\
NGC\,2004-082 & B1.5 V      &25700 &4.10 & 161      & A  & 4.42 & 13  &       \\
NGC\,2004-083 & B1.5:e      &	Be &  -- & 194      & -- & 4.42 & 13  & Binary\\
NGC\,2004-084 & B1.5 III    &27395 &4.00 &  36      & T06& 4.46 & 14  &       \\
NGC\,2004-085 & B2.5 III    &20450 &3.40 & 150      & A  & 4.18 & 10  &       \\
NGC\,2004-086 & B2 III      &21700 &3.85 &  14      & A  & 4.24 & 11  &       \\
NGC\,2004-087 & B1.5 V      &25700 &4.40 &  35      & A  & 4.40 & 13  &       \\
NGC\,2004-088 & B2.5 III    &20450 &3.65 & 200:     & A  & 4.18 & 10  & Binary\\
NGC\,2004-089 & B2.5e	    &	Be &  -- & 288      & -- & 4.31 & 12  &       \\
NGC\,2004-090 & O9.5 III    &32500 &4.10 &  16      & T06& 4.64 & 17  &       \\
NGC\,2004-091 & B1.5 III    &26520 &4.05 &  40      & T06& 4.42 & 13  & Binary\\
NGC\,2004-092 & B2e	    &	Be &  -- & 171      & -- & 4.34 & 12  &       \\
NGC\,2004-093 & B3 III      &20000 &3.65:& 189:     & He & 4.14 & 10  & Binary\\
NGC\,2004-094 & B2.5 III    &20450 &3.40 &  84:     & A  & 4.16 & 10  & Binary\\
NGC\,2004-095 & B1.5 V      &25700 &4.10 & 138      & A  & 4.38 & 13  &       \\
NGC\,2004-096 & B1.5e	    &	Be &  -- & 245      & -- & 4.38 & 13  &       \\
NGC\,2004-097 & B2 III      &21700 &3.60 &  62      & A  & 4.20 & 10  &       \\
NGC\,2004-098 & B2 III      &21700 &3.75 &  90      & A  & 4.20 & 10  &       \\
NGC\,2004-099 & B2 III      &21700 &3.40 & 119      & A  & 4.19 & 10  &       \\
NGC\,2004-100 & B1 Vn	    &26800 &3.70 & 323      & A  & 4.38 & 13  &       \\
NGC\,2004-101 & B2 III      &21700 &3.45 & 131      & A  & 4.18 & 10  &       \\
NGC\,2004-102 & B2 III      & SB2? &  -- &  64:     & -- & 4.18 & 10  & Binary\\
NGC\,2004-103 & B2 III      &21700 &3.85 &  35      & A  & 4.18 & 10  &       \\
NGC\,2004-104 & B1.5 V      &25700 &3.90 & 274      & A  & 4.34 & 12  &       \\
NGC\,2004-105 & B1.5 V      &25700 &3.90 & 235      & A  & 4.34 & 12  &       \\
NGC\,2004-106 & B2 III      &21700 &3.50 &  41      & A  & 4.17 & 10  &       \\
NGC\,2004-107 & B0.5 V      &28500 &3.90 & 146      & He & 4.43 & 14  & Binary\\
NGC\,2004-108 & B2.5 III    &22600 &4.00 &  13      & T06& 4.21 & 10  & Binary\\
NGC\,2004-109 & B2.5 III    &20450 &3.50 &  41      & A  & 4.11 & 10  & Binary\\
NGC\,2004-110 & B2 III      &21700 &3.40 & 121      & A  & 4.17 & 10  &       \\
NGC\,2004-111 & B2.5 III    &20450 &3.30 &  55      & A  & 4.08 &  9  &       \\
NGC\,2004-112 & B2 III      &21700 &3.65 &  72      & A  & 4.14 & 10  &       \\
NGC\,2004-113 & B2.5 IIIn   &20450 &3.25 & 307:     & A  & 4.08 &  9  &       \\
NGC\,2004-114 & B2 III      &21700 &3.60 &  59      & A  & 4.12 & 10  &       \\
NGC\,2004-115 & B2e	    &	Be &  -- &   8      & -- & 4.24 & 11  & Binary\\
NGC\,2004-116 & B2 III      &21700 &3.55 &  63      & A  & 4.12 & 10  &       \\
NGC\,2004-117 & B2 III      &21700 &3.60 &  75      & A  & 4.11 & 10  &       \\
NGC\,2004-118 & B1.5 V	    & SB2? & -- -&  90:     & -- & 4.25 & 12  & Binary\\
NGC\,2004-119 & B2 III	    &23210 &3.75 &  15      & T06& 4.15 & 10  &       \\
\\
\hline
\end{tabular}
\end{table*}

\end{document}